\documentclass[a4paper,12pt]{article}

\usepackage{amsmath}
\usepackage{amssymb}
\usepackage{slashed}
\usepackage{mathrsfs}
\usepackage{bbm}

\numberwithin{equation}{section}

\begin{document}

\begin{titlepage}
\begin{center}
\vspace*{-1.0cm}
\hfill DMUS--MP--14/14 \\

\vspace{2.0cm} {\Large \bf Supersymmetry of AdS and flat IIB backgrounds} \\[.2cm]

\vskip 2cm
S. W. Beck$^1$,  J. B.  Gutowski$^2$ and G. Papadopoulos$^1$
\\
\vskip .6cm

\begin{small}
$^1$\textit{  Department of Mathematics, King's College London
\\
Strand, London WC2R 2LS, UK.
\\
E-mails: samuel.beck@kcl.ac.uk,
\\
george.papadopoulos@kcl.ac.uk}
\end{small}\\*[.6cm]

\begin{small}
$^2$\textit{Department of Mathematics,
University of Surrey \\
Guildford, GU2 7XH, UK \\
Email: j.gutowski@surrey.ac.uk}
\end{small}\\*[.6cm]

\end{center}

\vskip 3.5 cm
\begin{abstract}

We present a systematic description of all warped  $AdS_n\times_w M^{10-n}$ and ${\mathbb{R}}^{n-1,1}\times_w M^{10-n}$  IIB backgrounds and
 identify the a priori number of supersymmetries $N$ preserved by these solutions. In particular, we find that the $AdS_n$ backgrounds preserve
$N=2^{[{n\over2}]} k$ for $n\leq 4$ and $N=2^{[{n\over2}]+1} k$ for $4<n\leq 6$ supersymmetries and for $k\in {\mathbb{N}}_{+}$ suitably restricted.  In addition under some  assumptions required for the applicability of the maximum principle, we demonstrate that the Killing spinors of $AdS_n$ backgrounds can be identified with the zero modes of  Dirac-like operators on $M^{10-n}$ establishing a new class of Lichnerowicz type theorems.  Furthermore, we  adapt some of these results to ${\mathbb{R}}^{n-1,1}\times_w M^{10-n}$ backgrounds with fluxes by taking the AdS radius to infinity. We find that these backgrounds
preserve $N=2^{[{n\over2}]} k$ for $2<n\leq 4$ and $N=2^{[{n+1\over2}]} k$ for $4<n\leq 7$ supersymmetries. We also demonstrate that the Killing spinors of $AdS_n\times_w M^{10-n}$
do not factorize into Killing spinors on $AdS_n$ and Killing spinors on $M^{10-n}$.

\end{abstract}

\end{titlepage}

\section{Introduction}

In the past thirty years, AdS backgrounds have found widespread applications in supergravity, string theory and M-theory. Following the original work of Freund and Rubin\cite{FR},
AdS backgrounds have been used in supergravity  compactifications, for  reviews see \cite{duff, grana} and references within,  and more recently in AdS/CFT \cite{maldacena}.
In particular, IIB AdS backgrounds, like $AdS_5\times S^5$, have been instrumental in the foundation  of AdS/CFT correspondence. Because of this, there is an extensive
literature in constructing such IIB backgrounds and in exploring their applications, for some selected publications see \cite{romans}-\cite{nunez}.  So far the construction of  most  supersymmetric AdS backgrounds
has been based on ansatzes  on either the form
of the fields, or of the Killing spinors\footnote{ As in  M-theory, the Killing spinors
do not factorize into a product of a Killing spinor on AdS and a Killing spinor on the transverse space. 
}.  As a result,  most of the investigations  have not been systematic, and to our knowledge
there is no full classification of AdS backgrounds.

In this paper, we initiate the classification of all IIB AdS backgrounds by specifying the fractions of supersymmetry preserved by such backgrounds. In a
future publication, we shall present their geometry \cite{adsII}. In particular, we shall solve the KSEs of IIB supergravity without any additional
assumptions\footnote{In the investigation that follows, we consider  backgrounds  up to discrete identifications.} on the fields and Killing spinors, apart from imposing on the former the symmetries of the AdS spaces. As a result, we identify
the a priori number of supersymmetries $N$ preserved by these backgrounds. In particular, we show that for $AdS_n\times_w M^{10-n}$,
\begin{eqnarray}
N=2^{[{n\over2}]} \, k~,~~~2\leq n\leq 4~;~~~N=2^{[{n\over2}]+1} \, k~,~~~4<n\leq 6~,
\label{susyc}
\end{eqnarray}
where $k\in {\mathbb{N}}_+$.
To prove the above result for $AdS_2$ backgrounds, we suitably restrict the transverse space $M^8$, eg $M^8$ can be taken to be compact without boundary, but such an assumption
is not necessary for the rest of the backgrounds.  Because of the classification results of \cite{n32, n31, n29, n28}, the number of supersymmetries $N$ are further restricted.  In particular, it is known that
there are no $AdS_2$ backgrounds with $N>26$ supersymmetries.  As a result $k<14$ in this case.  Similar restrictions apply to the other cases and the collected results
can be found in table 1. Furthermore all solutions preserving more than 16 supersymmetries are homogenous \cite{jose}.

Furthermore, we demonstrate that the the Killing spinors of $AdS_n\times_w M^{10-n}$ backgrounds can be identified with the zero modes of Dirac-like
operators on  $M^{10-n}$ coupled to fluxes.
For this and under suitable assumptions on $M^{10-n}$, we prove new Lichnerowicz type theorems which give a 1-1 correspondence between the solutions
of the KSEs and the zero modes of appropriate Dirac-like operators  ${\mathscr {D}}{}^{(\pm)}$.
As a consequence, we find that for $AdS_n$ backgrounds
\begin{eqnarray}
N=2 (N_-+ \mathrm{Index} (D))~,~~~n=2; ~~~~N=\ell(n) N_-~,~~~n>2~,
\label{indexf}
\end{eqnarray}
 where $D$ is  the Dirac operator possibly twisted with a $U(1))$ bundle on $M^8$,  $N_-=\mathrm{dim}\, \mathrm{Ker} {\mathscr {D}}{}^{(-)}$ and $\ell(n)=2^{[{n\over2}]}$ for $2< n\leq 4$ and $\ell(n)=4$
for $4<n\leq 6$. Observe that $N_-$ is even for $n=5$ and $N_-=4 k$ for $n=6$.

Our $AdS_n$ results can be adapted to  ${\mathbb{R}}^{n-1,1}\times_w M^{10-n}$ backgrounds in the limit that the $AdS$ radius goes to infinity. This limit is smooth
in all our local computations but some of the regularity assumptions needed to establish some of our global results, like the new Lichnerowicz type theorems,
are no longer  valid. Nevertheless, we have solved the   KSEs  and the number of supersymmetries preserved by the ${\mathbb{R}}^{n-1,1}\times_w M^{10-n}$  is
\begin{eqnarray}
N= 2^{[{n\over 2}]} k~,~~~2<n\leq 4~;~~~N= 2^{[{n+1\over 2}]} k~,~~~4<n\leq 8~.
\label{susycf}
\end{eqnarray}
The supersymmetries  preserved by ${\mathbb{R}}^{1,1}\times_w M^{8}$ backgrounds cannot be decided.  This is because to show that $AdS_2\times_w M^8$
preserves an even number of supersymmetries requires the use of a maximum principle argument which may not be valid  for ${\mathbb{R}}^{1,1}\times_w M^{8}$ backgrounds.
The results
have been tabulated in table 2.

To prove our results, we have to solve the gravitino and dilatino KSEs of IIB supergravity for AdS backgrounds. For this, we have used the observation in \cite{adshor} that all AdS backgrounds can be described as near horizon geometries of extreme Killing horizons. This facilitates
the integrability of the KSEs along the AdS directions. First after decomposing the Killing spinor as $\epsilon=\epsilon_++\epsilon_-$, where $\Gamma_\pm \epsilon_\pm=0$ are lightcone projections,
we  use the near horizon results of \cite{iibhor} to integrate
the KSEs along two lightcone directions, see also \cite{5dindex, JGGP1}. After the integration the Killing spinors are written as $\epsilon_\pm=\epsilon_\pm(\eta_+, \eta_-; r,u)$,  where $(r,u)$ are appropriate
 coordinates, $\epsilon_\pm\vert_{r=u=0}=\eta_\pm$ and $\eta_\pm$ are spinors which are now localized on the co-dimension two subspace  ${\cal S}$ given by $r=u=0$ which is the horizon spatial section.  To identify the remaining
 independent KSEs we use a key result of \cite{iibhor}, in which it is shown, after a rather involved argument
 and the use of field equations,  that the remaining independent KSEs are derived from the
naive restriction of the original KSEs on ${\cal S}$. The final result is two sets of KSEs on ${\cal S}$ one acting on $\eta_+$ and the other on $\eta_-$. Each set contains a parallel transport equation associated to the original gravitino
 and one algebraic KSE associated to the original dilatino KSE. This suffices to integrate the KSEs on $AdS_2\times_w M^8$ along the  $AdS_2$ directions as  $M^8={\cal S}$.

For the rest of the $AdS_n$ backgrounds, the KSEs can be integrated along all AdS directions. For this, the Killing spinors $\eta_\pm$ are expressed as $\eta_+=\eta_+(\sigma_+, \tau_+, x)$ and $\eta_-=\eta_-(\sigma_-, \tau_-, x)$
   where now $\sigma_\pm$ and $\tau_\pm$ are localized on $M^{10-n}$ and $x$ denotes AdS coordinates. The independent KSEs can be organized into four sets of equations on $M^{10-n}$, one for each $\sigma_\pm$ and $\tau_\pm$.
   Each set contains three KSEs. The first two are associated to the original gravitino and dilatino KSEs, and there is an additional algebraic KSE which arises
   from the process of integrating over the remaining $AdS_n$ directions.

The counting of supersymmetries proceeds as follows. The proof that for $AdS_2$ backgrounds the number of supersymmetries is even and the formula
in (\ref{indexf}) follows from the results of \cite{iibhor} on the number of supersymmetries preserved by near horizon geometries. For this, it is
required that the fields and $M^8$ satisfy the conditions for the Hopf maximum principle to apply. In particular, $M^8$ can be taken to be compact and connected without boundary, and the fields smooth.

The counting of supersymmetries for the remaining $AdS_n$, $n>2$, backgrounds is done in a different way. In particular it is observed that
there are Clifford algebra operators which intertwine between the four sets of KSEs. So given a solution in one set of KSEs, one automatically
has solutions in the other sets. After identifying all the intertwining Clifford algebra operators for each of the $AdS_n$ backgrounds, one derives the results of (\ref{susyc}). It should be noted that for the proof of (\ref{susyc}), there is no need to put any restriction on $M^{10-n}$ or on the fields as is done for $AdS_2$ case.

The proof of the formula for $N$ in  (\ref{indexf}) for $AdS_2$ backgrounds is similar to that for near horizon geometries. As has been established in \cite{iibhor},
 this necessitates
the proof of two Lichnerowicz type theorems, one for each $\eta_\pm$ spinor. The proof of these theorems are based on the Hopf maximum principle and a partial
integration formula. A simplification for $AdS_2$ backgrounds is that both Lichnerowicz type theorems can be demonstrated using only the Hopf maximum principle.

Furthermore, the proof of the formula for $N$ in (\ref{indexf}) for the remaining $AdS_n$ backgrounds again requires the proof of four Lichnerowicz type theorems
one for each of the spinors $\sigma_\pm$ and $\tau_\pm$.  The proof of these theorems utilizes the Hopf maximum principle on the square of the length of the spinors
$\sigma_\pm$ and $\tau_\pm$. Instrumental in the proof is the use of the field equations and the choice of  modified  Dirac-like operators  ${\mathscr {D}}{}^{(\pm)}$ on $M^{10-n}$
coupled to fluxes. These modified  Dirac-like operators are constructed as an appropriate linear combination of the  Dirac operator associated with the gravitino KSEs on $M^{10-n}$
and the new algebraic KSEs that arise in the analysis. It is remarkable that the zero modes of these modified  Dirac-like operators  solve not only the parallel transport equation, which is expected
from the classic Lichnerowicz theorem, but also solve the two algebraic KSEs.  Note that unlike the $AdS_2$ case, in the remaining $AdS_n$ cases, there is no contribution
from the index of the Dirac operator to $N$, as it vanishes.

To prove the formula (\ref{susycf}) for the number of supersymmetries preserved by ${\mathbb{R}}^{n-1,1}\times_w M^{10-n}$ backgrounds, one takes the  AdS radius  to infinity.
All the local computations for $AdS_n$ backgrounds are valid in this limit and so carry through without alterations. However, as we shall explain this is not
the case for some of the global results, like the Lichnerowicz type theorems, which require for their validity certain regularity restrictions on the fields
which are no longer  valid. Another significant difference, which affects the counting of supersymmetries, is that in the limit of infinite AdS radius the spinors
$\sigma_\pm$ and $\tau_\pm$ are no longer  linearly independent. Because of this, the counting of supersymmetries between $AdS_n$ and ${\mathbb{R}}^{n-1,1}$
backgrounds is different.

This paper is organized as follows. In section 2, we explain how the $AdS_n\times_w M^{10-n}$ backgrounds can be written as near horizon geometries, and summarize
the key results of \cite{iibhor} regarding the solution of IIB KSEs for near horizon geometries. In section 3, we solve the KSEs for $AdS_2\times_w M^{8}$
backgrounds and find under which conditions the number of supersymmetries preserved is even. In section 4, we prove  new  Lichnerowicz type theorems for
$AdS_2\times_w M^{8}$ backgrounds and identify Killing spinors with the zero modes of a Dirac-like  operator on $M^8$.  In section 5 and 6, 7 and 8, 9 and 10,
11 and 12, the KSEs are solved and identification of Killing spinors as zero modes of Dirac-like operators on $M^{10-n}$ is  done, for $AdS_3\times_w M^{7}$,
$AdS_4\times_w M^{6}$, $AdS_5\times_w M^{5}$ and $AdS_6\times_w M^{4}$ backgrounds, respectively, resulting in the proof of formulae (\ref{susyc}) and
(\ref{indexf}).  In section 13, we show that there are no supersymmetric $AdS_n\times_w M^{10-n}$ backgrounds for $n>6$.  In section 14, we
prove the formula (\ref{susycf}) for  ${\mathbb{R}}^{n-1,1}\times_w M^{10-n}$ backgrounds and in section 15, we give our conclusions.
In appendix A, we have summarized our conventions. In appendices B, C, D, E, we present the proof of the maximum principle formulae on the length of zero modes of
${\mathscr {D}}{}^{(\pm)}$,
required to prove the new Lichnerowicz type theorems for $AdS_n\times_w M^{10-n}$ backgrounds, $2<n<7$.

\section{AdS and near horizon geometries}

\subsection{Warped AdS and flat backgrounds}

The warped AdS and flat backgrounds can be written universally as near horizon geometries \cite{adshor}.  Let $F$, $G$ and $P$ be the 5-, 3- and 1-form field strengths of IIB supergravity. All  AdS backgrounds can be described in terms of the fields
\begin{eqnarray}
ds^2 &=&2 {\bf{e}}^+ {\bf{e}}^- + ds^2({\cal S})~,~~~F= r {\bf{e}}^+ \wedge X + {\bf{e}}^+ \wedge {\bf{e}}^- \wedge Y + \star_8 Y~,~~
\cr
G &=& r {\bf{e}}^+ \wedge L + {\bf{e}}^+ \wedge {\bf{e}}^- \wedge \Phi + H~,~~~P = \xi~,
\label{hormetr}
\end{eqnarray}
where we have introduced the frame
\begin{eqnarray}
\label{basis1}
{\bf{e}}^+ = du, \qquad {\bf{e}}^- = dr + rh -{1 \over 2} r^2 \Delta du, \qquad {\bf{e}}^i = e^i_I dy^I~,
\end{eqnarray}
and
\begin{eqnarray}
ds^2({\cal S})=\delta_{ij} {\bf{e}}^i {\bf{e}}^j~,
\end{eqnarray}
is the metric on the horizon spatial section ${\cal S}$ which is the co-dimension 2 submanifold given by the equations $r=u=0$. In addition,
  the self-duality of $F$ requires that  $X = - \star_8 X$.
The dependence on the coordinates $u$ and $r$ is explicitly given. $\Delta$, $h$, $Y$ are 0-, 1- and 3-forms
on ${\cal S}$, respectively, $\Phi$, $L$ and $H$ are $\lambda$-twisted 1-, 2- and 3-forms on ${\cal S}$, respectively, and $\xi$ is a $\lambda^2$-twisted 1-form on ${\cal S}$, where $\lambda$ arises
 from the pull back of the canonical bundle of the scalar manifold\footnote{The scalar manifold can also be taken as the fundamental domain of the modular group but we shall not dwell on this.} $SU(1,1)/U(1)$ on ${\cal S}$.  Furthermore, the Bianchi identities imply that
 \begin{eqnarray}
X = d_hY -{i \over 8} (\Phi \wedge {\bar{H}} - {\bar{\Phi}} \wedge H)~,~~~L = d_h \Phi  -i \Lambda \wedge \Phi + \xi \wedge {\bar{\Phi}}~,
\end{eqnarray}
and so $X$ and $L$ are not independent fields.

Moreover, viewing the backgrounds  $AdS_n\times_w M^{10-n}$ as near horizon geometries, the spatial horizon sections ${\cal S}$ are ${\cal S}=H^{n-2}\times_w M^{10-n}$, ie  warped products of hyperbolic (n-2)-dimensional space with $M^{10-n}$. This can be easily seen after the fields are stated explicitly for each case below.

Although all AdS backgrounds are described by (\ref{hormetr}), the field dependence of individual AdS cases differs. To address this, we shall separately state the fields
 in each case as follows.

\subsubsection{$AdS_2\times_w M^8$}

In this case $M^8={\cal S}$ and the fields become
\begin{eqnarray}
ds^2 &=& 2 du (dr + r h - {1 \over 2} r^2 \Delta du)+ ds^2(M^9)~,~~~F=   {\bf{e}}^+ \wedge {\bf{e}}^- \wedge Y + \star_8 Y~,~~
\cr
G &=&   {\bf{e}}^+ \wedge {\bf{e}}^- \wedge \Phi + H~,~~~P = \xi~,
\end{eqnarray}
where
\begin{eqnarray}
h=-2 A^{-1} dA=\Delta^{-1} d\Delta ~,~~~X=L=0~.
\end{eqnarray}
 Observe that $dh=0$ and $A$ is the warp factor.

 \subsubsection{$AdS_3\times_w M^7$}
 The fields are
 \begin{eqnarray}
ds^2 &=& 2 du (dr + r h )+ A^2 dz^2 +ds^2(M^7)~,~~~F=  A {\bf{e}}^+ \wedge {\bf{e}}^- \wedge dz\wedge Y - \star_7 Y
\cr
G &=& A  {\bf{e}}^+ \wedge {\bf{e}}^- \wedge dz \wedge \Phi + H~,~~~P = \xi~,
 \end{eqnarray}
 where
 \begin{eqnarray}
h = -{2 \over \ell}dz -2 A^{-1} dA,~~~\Delta=0~,~~~~X=L=0~,
\end{eqnarray}
and $\ell$ is the radius of $AdS$.

 \subsubsection{$AdS_4\times_w M^6$}
 The fields are
 \begin{eqnarray}
ds^2 &=& 2 du (dr + r h )+ A^2 (dz^2+ e^{2z/\ell} dx^2) +ds^2(M^7)~,
\cr
F &=&A^2 e^{z/\ell}\, {\bf{e}}^+ \wedge {\bf{e}}^- \wedge dz \wedge dx \wedge Y+\star_6 Y~,
 \cr
G &=&   H~,~~~P = \xi~,
 \end{eqnarray}
 where
 \begin{eqnarray}
h = -{2 \over \ell}dz -2 A^{-1} dA,~~~\Delta=0~,~~~~X=L=0~.
\end{eqnarray}

 \subsubsection{$AdS_5\times_w M^5$}
The fields are
\begin{eqnarray}
ds^2 &=& 2 du (dr + r h )+ A^2 (dz^2+ e^{2z/\ell} (dx^2+dy^2) +ds^2(M^{5})~,~~~
 \cr
F &=& Y \left[ A^3 e^{2z/\ell}{\bf{e}}^+ \wedge {\bf{e}}^- \wedge dz \wedge dx \wedge dy - \operatorname{dvol} \left(M^5\right) \right]~,
\cr
G&=&H~,~~~P = \xi~,
 \end{eqnarray}
 where
 \begin{eqnarray}
h = -{2 \over \ell}dz -2 A^{-1} dA,~~~\Delta=0~,~~~~X=L=0~.
\end{eqnarray}

 \subsubsection{$AdS_6\times_w M^4$}

 The fields are
 \begin{eqnarray}
ds^2 &=& 2 du (dr + r h )+ A^2 \big(dz^2+ e^{2z/\ell} \sum_{a=1}^3(dx^a)^2\big) +ds^2(M^{4})~,~~~F = 0  ,
 \cr
G &=&   H~,~~~P = \xi~,
 \end{eqnarray}
 where
 \begin{eqnarray}
h = -{2 \over \ell}dz -2 A^{-1} dA,~~~\Delta=0~,~~~~X=L=0~.
\end{eqnarray}

It should be noted that the warped backgrounds ${\mathbb{R}}^{n-1,1}\times_w M^{10-n}$ are included in our analysis.  They arise in the limit
that the AdS radius $\ell$ goes to infinity.  This limit is smooth for all our field configurations presented above. However,
some statements that apply for $AdS$  do not extend to the flat backgrounds.  Because of this some care must be taken when adapting
the results we obtain for $AdS$ backgrounds to the limit of infinite radius.

\subsection{Bianchi identities and field equations}
\label{horbf}

It is clear from the expressions of the fields for the AdS backgrounds in the previous section that $L=X=0$ and $dh=0$.  As  a result, we have
\begin{eqnarray}
d_hY -{i \over 8} (\Phi \wedge {\bar{H}} - {\bar{\Phi}} \wedge H)=0,~~~ d_h \Phi  -i \Lambda \wedge \Phi + \xi \wedge {\bar{\Phi}}=0~.
\label{ba0}
\end{eqnarray}
Furthermore, the
remaining Bianchi identities for the backgrounds (\ref{hormetr}) are
\begin{eqnarray}
d \star_8 Y &=& {i \over 8} H \wedge {\bar{H}}~,~~~dH = i \Lambda \wedge H - \xi \wedge {\bar{H}}~, ~~~
\cr
d \xi &=& 2i \Lambda \wedge \xi~,~~~d \Lambda = -i \xi \wedge {\bar{\xi}}~,
\label{ba}
\end{eqnarray}
where $\Lambda$ is a $U(1)$ connection of $\lambda$, see \cite{iibhor} for more details.

The independent field equations of the AdS backgrounds (\ref{hormetr})  are
\begin{eqnarray}
\label{feq1}
{\tilde{\nabla}}^i \Phi_i -i \Lambda^i \Phi_i - \xi^i {\bar{\Phi}}_i +{2 i \over 3} Y_{\ell_1 \ell_2 \ell_3} H^{\ell_1 \ell_2 \ell_3}=0~,
\end{eqnarray}

\begin{eqnarray}
\label{feq3}
{\tilde{\nabla}}^\ell H_{\ell ij} -i \Lambda^\ell H_{\ell ij}- h^\ell H_{\ell ij}  - \xi^\ell {\bar{H}}_{\ell ij}
+{2i \over 3}( \star_8 Y_{ij \ell_1 \ell_2 \ell_3} H^{\ell_1 \ell_2 \ell_3} - 6 Y_{ij \ell} \Phi^\ell) =0~,
\nonumber \\
\end{eqnarray}

\begin{eqnarray}
\label{feq4}
{\tilde{\nabla}}^i \xi_i -2i \Lambda^i \xi_i - h^i \xi_i +{1 \over 24}(-6 \Phi^2 + H^2) =0~,
\end{eqnarray}

\begin{eqnarray}
\label{feq5}
{1 \over 2} {\tilde{\nabla}}^i h_i - \Delta - {1 \over 2} h^2 + {2 \over 3} Y^2
+{3 \over 8}  \Phi^i {\bar{\Phi}}_i +{1 \over 48}\parallel H \parallel^2 =0~,
\end{eqnarray}
and
\begin{eqnarray}
\label{feq7}
{\tilde{R}}_{ij} + {\tilde{\nabla}}_{(i} h_{j)} -{1 \over 2} h_i h_j +4 Y^2_{ij}
+{1 \over 2} \Phi_{(i} {\bar{\Phi}}_{j)} -2 \xi_{(i} {\bar{\xi}}_{j)}
-{1 \over 4} H_{\ell_1 \ell_2 (i} {\bar{H}}_{j)}{}^{\ell_1 \ell_2}
\nonumber \\
+ \delta_{ij} \bigg(-{1 \over 8} \Phi_\ell {\bar{\Phi}}^\ell -{2 \over 3} Y^2
+{1 \over 48}\parallel H \parallel^2 \bigg) =0~,
\end{eqnarray}
where $\tilde \nabla$ and $\tilde R$ are the Levi-Civita connection and the Ricci tensor of ${\cal S}$, respectively.
There is an additional field equation which is not independent because they follow from those above. This is
\begin{eqnarray}
{1 \over 2} {\tilde{\nabla}}^2 \Delta -{3 \over 2} h^i {\tilde{\nabla}}_i \Delta -{1 \over 2} \Delta {\tilde{\nabla}}^i h_i + \Delta h^2
 =0~,
\nonumber \\
\label{feq8}
\end{eqnarray}
which we state because it  is useful in the investigation of the KSEs.

\subsection{Killing spinor equations} \label{lightcone}

The gravitino and dilatino KSEs of IIB supergravity \cite{schwarz1, schwarz2}  are
\begin{eqnarray}
\label{gkse}
&&\bigg(\nabla_M -{i \over 2} Q_M
+{i \over 48} \slashed{F}_{M } \bigg) \epsilon
-{1 \over 96} \bigg(\Gamma\slashed{G}_M  -9 \slashed{G}_{M } \bigg) C* \epsilon
=0~,
\end{eqnarray}
\begin{eqnarray}
\label{akse}
\slashed{P} C* \epsilon +{1 \over 24} \slashed{G} \epsilon =0~,
\end{eqnarray}
respectively, where $Q$ is a $U(1)$ connection of $\lambda$.

These KSEs can be solved for the fields (\ref{hormetr}) along the directions $u,r$.  For this first decompose $\epsilon=\epsilon_++\epsilon_-$, where $\Gamma_\pm\epsilon_\pm=0$.
Then a direct substitution into (\ref{gkse}) and (\ref{akse}) reveals that the Killing spinor can be expressed as
\begin{eqnarray}
\label{st1}
\epsilon_+ = \phi_+,~~
\epsilon_-=\phi_- + r \Gamma_- \Theta_+ \phi_+~;~~\phi_+ =\eta_+ + u \Gamma_+ \Theta_-\eta_-~,~~\phi_- =\eta_-~,
\end{eqnarray}
where
$\eta_\pm$ do not depend on both $u$ and $r$ coordinates and
\begin{eqnarray}
\Theta_\pm= \bigg({1 \over 4} \slashed{h}
\pm{i \over 12} \slashed {Y} \bigg)
+\bigg( {1 \over 96} \slashed {H} \pm {3 \over 16} \slashed{\Phi} \bigg) C*~.
\end{eqnarray}
After some extensive computation using the field equations described in \cite{iibhor}, one can show that the independent KSEs for the backgrounds (\ref{hormetr}) are
\begin{eqnarray}
\nabla_i^{(\pm)} \eta_\pm=0~,~~~{\cal A}^{(\pm)}\eta_\pm=0~,
\label{horkses}
\end{eqnarray}
where
\begin{eqnarray}
\label{skse1}
\nabla^{(\pm)}_i&\equiv&{\tilde{\nabla}}_i  + \bigg(-{i \over 2} \Lambda_i \mp{1 \over 4} h_i
\mp{i \over 4} \slashed{Y}_{i }
\pm{i \over 12}\Gamma\slashed{Y}_i \bigg)
\cr
&&+ \bigg(\pm{1 \over 16} \Gamma\slashed{\Phi}_i \mp{3 \over 16} \Phi_i
-{1 \over 96}\Gamma\slashed{H}_i
+{3 \over 32} \slashed{H}_{i} \bigg) C*~,
\end{eqnarray}
and
\begin{eqnarray}
\label{alg3}
{\cal A}^{(\pm)}\equiv\bigg(\mp{1 \over 4}\slashed{\Phi} +{1 \over 24} \slashed{H}
\bigg)  + \slashed{\xi} C*  ~.
\end{eqnarray}
It turns out that (\ref{horkses}) are the restriction of (\ref{gkse}) and (\ref{akse}) on the horizon section ${\cal S}$ for $\epsilon$ given in (\ref{st1}).

Furthermore, one can show that if $\eta_-$ is a solution to the KSEs, then
\begin{eqnarray}
\eta_+=\Gamma_+\Theta_-\eta_-~,
\end{eqnarray}
also solves the KSEs. This is the first indication that IIB horizons exhibit supersymmetry enhancement. Indeed if ${\cal S}$ is compact and the fluxes
do not vanish, one can show \cite{iibhor} that $\mathrm{Ker}\Theta_-=\{0\}$ and so $\eta_+$ given in the above equation yields an additional supersymmetry.

Although the following integrability conditions
\begin{eqnarray}
\label{alg1}
\big({1 \over 2} \Delta
+2 \Theta_- \Theta_+ \big)\eta_+
=0~,~~~
\big({1 \over 2} \Delta
+2 \Theta_+ \Theta_- \big)  \eta_- =0~,
\end{eqnarray}
are implied  from the (\ref{horkses}) KSEs, the field equations and the Bianchi identities  it is convenient for the analysis that follows to include them.  As we shall see, they are instrumental in the solution
of the KSEs along the $AdS_n$ directions for $n>2$.

\subsection{Horizon Dirac equations}

Before we complete this section, we shall summarize the results of \cite{iibhor} on the relation between Killing spinors
 and zero modes of Dirac-like operators for IIB horizons. We have seen that the gravitino KSE gives rise to two parallel transport equations on ${\cal S}$ associated with the covariant derivatives
$\nabla^{(\pm)}$  (\ref{skse1}). If $S^\pm$ are the complex chiral spin bundles over ${\cal S}$, then
$\nabla^{\pm}: \Gamma(S^\pm\otimes \lambda^{1\over2})\rightarrow \Gamma(\Lambda^1({\cal S})\otimes S^\pm\otimes \lambda^{1\over2})$,
where $\Gamma(S^\pm\otimes \lambda^{1\over2})$ are the smooth sections of $S^\pm\otimes \lambda^{1\over2}$. In turn,
one can define the associated horizon Dirac operators
\begin{eqnarray}
{\cal D}^{(\pm)}\equiv \Gamma^i \nabla^{(\pm)}_i=\Gamma^i \tilde\nabla_i+\Psi^\pm~,
\label{dirac}
\end{eqnarray}
where
\begin{eqnarray}
\Psi^\pm\equiv \Gamma^i \Psi_i^{(\pm)}=-{i \over 2} \slashed{\Lambda} \mp {1 \over 4} \slashed{h} \pm {i \over 6} \slashed{Y}+ \big( \pm {1 \over 4} \slashed{\Phi} +{1 \over 24} \slashed{H}
\big) C* \ .
\nonumber \\
\end{eqnarray}
Clearly the $\nabla^{\pm}$ parallel spinors   are  zero modes of ${\cal D}^{(\pm)}$. For ${\cal S}$ compact, one can also prove the converse, ie that
all zero modes of the horizon Dirac equations ${\cal D}^{(\pm)}$ are Killing spinors. Therefore, one can
 establish
\begin{eqnarray}
\nabla^{(\pm)}\eta_\pm=0~,~~~{\cal A}^{(\pm)}\eta_\pm=0\Longleftrightarrow {\cal D}^{(\pm)}\eta_\pm=0~.
\label{lich}
\end{eqnarray}
The proof of the above statement for $\eta_+$ spinors utilizes the Hopf maximum principle on $\parallel \eta_+\parallel^2$ while for $\eta_-$ it employs a partial
integration formula.  In the former case, one also finds that $\parallel \eta_+\parallel = \mathrm{const}$. Similar theorems have been proven for other theories
in \cite{5dindex, JGGP1}.

\section{$AdS_2$: Local analysis}

\subsection{Fields, Bianchi identities and field equations}

For $AdS_2$ backgrounds $M^8={\cal S}$ and the fields on ${\cal S}$ are
\begin{eqnarray}
&&ds^2({\cal S})=ds^2(M^8)~,~~~\tilde F^3=Y~,~~~\tilde F^5=\star_8 Y~,~~~
\cr
&&\tilde G^1=\Phi~,~~~\tilde G^3=H~,~~~\tilde P=\xi~.
\end{eqnarray}
Next, we set
\begin{eqnarray}
\Delta=\ell^{-2} A^{-2}~,
\end{eqnarray}
which satisfies (\ref{feq8}), where $\ell$ is the radius of $AdS_2$. Using these,  the Bianchi identities (\ref{ba0}) and (\ref{ba})   can now be written as
\begin{eqnarray}
d(A^{-2} Y) -{i A^{-2}\over 8} (\Phi \wedge {\bar{H}} - {\bar{\Phi}} \wedge H)=0~,
\cr
 d(A^{-2} \Phi)  -i A^{-2} \Lambda \wedge \Phi + A^{-2}\xi \wedge {\bar{\Phi}}=0~,
\end{eqnarray}
and
\begin{eqnarray}
d \star_8 Y &=& {i \over 8} H \wedge {\bar{H}}~,~~~dH = i \Lambda \wedge H - \xi \wedge {\bar{H}}~, ~~~
\cr
d \xi &=& 2i \Lambda \wedge \xi~,~~~d \Lambda = -i \xi \wedge {\bar{\xi}}~,
\label{ads2ba}
\end{eqnarray}
respectively, where $\Lambda$ is a $U(1)$ connection of $\lambda$ restricted on ${\cal S}$.

Similarly, the field equations read as
\begin{eqnarray}
\label{ads2feq1}
{\tilde{\nabla}}^i \Phi_i -i \Lambda^i \Phi_i - \xi^i {\bar{\Phi}}_i +{2 i \over 3} Y_{\ell_1 \ell_2 \ell_3} H^{\ell_1 \ell_2 \ell_3}=0~,
\end{eqnarray}

\begin{eqnarray}
\label{ads2feq3}
{\tilde{\nabla}}^\ell H_{\ell ij} -i \Lambda^\ell H_{\ell ij}+2 A^{-1} \partial^\ell A H_{\ell ij}  - \xi^\ell {\bar{H}}_{\ell ij}
+{2i \over 3}( \star_8 Y_{ij \ell_1 \ell_2 \ell_3} H^{\ell_1 \ell_2 \ell_3} - 6 Y_{ij \ell} \Phi^\ell) =0~,
\end{eqnarray}

\begin{eqnarray}
\label{ads2feq4}
{\tilde{\nabla}}^i \xi_i -2i \Lambda^i \xi_i +2 A^{-1} \partial^i A \xi_i +{1 \over 24}(-6 \Phi^2+ H^2) =0~,
\end{eqnarray}

\begin{eqnarray}
\label{ads2feq5}
-A^{-1}\tilde \nabla^2 A -A^{-2} \partial^iA \partial_i A -\ell^{-2} A^{-2}+ {2 \over 3} Y^2
+{3 \over 8} \Phi^i {\bar{\Phi}}_i +{1 \over 48} \parallel H\parallel^2 =0~,
\end{eqnarray}
and
\begin{eqnarray}
\label{ads2feq7}
{R}^{(8)}_{ij}-2 A^{-1} \tilde\nabla_i\partial_j A +4 Y^2_{ij}
+{1 \over 2} \Phi_{(i} {\bar{\Phi}}_{j)} -2 \xi_{(i} {\bar{\xi}}_{j)}
-{1 \over 4} H_{\ell_1 \ell_2 (i} {\bar{H}}_{j)}{}^{\ell_1 \ell_2}
\nonumber \\
+ \delta_{ij} \bigg(-{1 \over 8} \Phi_\ell {\bar{\Phi}}^\ell -{2 \over 3} Y^2
+{1 \over 48} \parallel H\parallel^2 \bigg) =0~,
\end{eqnarray}
where $ R^{(8)}$ is the Ricci tensor of ${\cal S}=M^8$.

\subsubsection{The warp factor $A$ is no-where vanishing}

To see this, assume that $A$ is not identically zero. Thus there is a point in $M^8$ such that $A\not=0$. Multiplying (\ref{ads2feq5}) with $A^2$ evaluated
at a point for which $A\not=0$, one finds
\begin{eqnarray}
-A\tilde \nabla^2 A -\partial^iA \partial_i A -\ell^{-2} + {2 \over 3} A^2 Y^2
+{3 \over 8} A^2 \Phi^i {\bar{\Phi}}_i
+{1 \over 48} A^2\parallel H\parallel^2  =0~,
\end{eqnarray}
Next taking a sequence that converges at a point where $A$ vanishes, one finds an inconsistency
as the term involving the AdS radius $\ell$ cannot vanish.  Therefore there are no smooth solutions for which $A$ vanishes at some point on the spacetime.
A more detailed argument for this has been presented in \cite{Mads}.

This property depends crucially on $\ell$ taking a finite value. In particular, it is not valid in the limit that $\ell$ goes to infinity, and so
one cannot conclude that $A$ is no-where vanishing for ${\mathbb{R}}^{n-1,1}$ backgrounds.

\subsection{Killing spinor equations}

The KSEs on ${\cal S}={\cal M}^8$ are

\begin{eqnarray}
\nabla_i^{(\pm)} \eta_\pm=0~,~~~{\cal A}^{(\pm)}\eta_\pm=0~,
\label{ads2kse}
\end{eqnarray}
where
\begin{eqnarray}
\label{skse1x}
\nabla^{(\pm)}_i&\equiv&{\tilde{\nabla}}_i  + \bigg(-{i \over 2} \Lambda_i \pm{1 \over 2} A^{-1} \partial_i A
\mp{i \over 4}\slashed {Y}_{i}
\pm{i \over 12} \Gamma\slashed{Y}_i  \bigg)
\cr
&&+ \bigg(\pm{1 \over 16} \Gamma\slashed{\Phi}_i  \mp{3 \over 16} \Phi_i
-{1 \over 96}  \Gamma\slashed{H}_i
+{3 \over 32} \slashed{H}_{i } \bigg) C*~,
\end{eqnarray}
and
\begin{eqnarray}
\label{alg3}
{\cal A}^{(\pm)}\equiv\bigg(\mp{1 \over 4} \slashed{\Phi} +{1 \over 24}\slashed{ H}
\bigg)  + \slashed{\xi}  C*  ~.
\end{eqnarray}

Furthermore, if $\eta_-$ is a Killing spinor, then
\begin{eqnarray}
\eta_+=\Gamma_+\Theta_-\eta_-~,
\end{eqnarray}
is also a Killing spinor, where now
\begin{eqnarray}
\Theta_\pm= \bigg(-{1 \over 4} \slashed{\partial}\log A^2
\pm{i \over 12} \slashed{Y}  \bigg)
+\bigg( {1 \over 96} \slashed {H} \pm {3 \over 16} \slashed{\Phi}\bigg) C*~.
\end{eqnarray}
It is not apparent that $\eta_+\not=0$ as $\eta_-$ may be in the kernel of $\Theta_-$. To establish under which conditions
$\eta_+\not=0$, one has to impose additional restrictions on $M^8$. However if $\eta_+\not=0$, then the solutions exhibit
supersymmetry enhancement.

\subsection{Counting supersymmetries}

The analysis so far is not sufficient to establish either the formula in (\ref{susyc}) or (\ref{indexf}) regarding the number of supersymmetries $N$
preserved by the $AdS_2$ backgrounds.  For this, some additional restrictions on $M^8$ are required.  We shall explore these
in the next section.

\section{$AdS_2$:  Global analysis}

The main results of this section are to demonstrate that under certain assumptions, there is a 1-1 correspondence between Killing spinors
and zero modes of Dirac operators on $M^8$ coupled to fluxes, and use this to count the supersymmetries $N$ of $AdS_2$ backgrounds. Given
the gravitino KSE in (\ref{ads2kse}) and in particular the (super)covariant
derivatives $\nabla^{(\pm)}$, one can construct the Dirac-like operators
\begin{eqnarray}
{\cal D}^{(\pm)}\equiv \Gamma^i \nabla^{(\pm)}_i=\Gamma^i \tilde\nabla_i+\Psi^\pm~,
\label{dirac}
\end{eqnarray}
on $M^8$,
where
\begin{eqnarray}
\Psi^\pm\equiv \Gamma^i \Psi_i^{(\pm)}=-{i \over 2} \slashed{\Lambda}  \pm {1 \over 4} \slashed{\partial}\log A^2 \pm {i \over 6} \slashed{Y}+ \big( \pm {1 \over 4} \slashed{\Phi}
 +{1 \over 24}\slashed{ H} \big) C*~.
\nonumber \\
\end{eqnarray}
Clearly all parallel spinors $\eta_\pm$, ie $\nabla^{(\pm)}\eta_\pm=0$,  are  zero modes of ${\cal D}^{(\pm)}$, ie ${\cal D}^{(\pm)}\eta_\pm=0$. The task  is to prove the converse.

\subsection{A Lichnerowicz type theorem for ${\cal D}^{(+)}$}

The proof of this converse is a  Lichnerowicz type theorem and the proof is similar to that given in \cite{iibhor} for horizon Dirac operators. Because of this, we shall
not give details of the proof. The novelty of
 this theorem is that the converse implies that the zero modes of ${\cal D}^{(+)}$ solve both the gravitino and dilatino KSEs. In particular, assuming that ${\cal D}^{(+)}\eta_+=0$ and after some algebra which involves the use of field equations, one can establish that
 \begin{eqnarray}
{\tilde{\nabla}}^i {\tilde{\nabla}}_i \parallel \eta_+\parallel^2
+ \partial^i\log A^2\, {\tilde{\nabla}}_i \parallel \eta_+\parallel^2
= 2 \parallel \nabla^{(+)}\eta_+ \parallel^2+  \parallel{\cal{A}}^{(+)}\eta_+\parallel^2~.
\label{lichident}
\end{eqnarray}
It is then a consequence of the maximum principle that the only solution of the above equation is $\parallel \eta_+\parallel=\mathrm{const}$ and that $\eta_+$ is a Killing spinor. In particular, this is the case provided  $M^8$ is compact and the fields are smooth.

\subsection{A Lichnerowicz type theorem for ${\cal D}^{(-)}$} \label{slich2}

The proof that the zero modes of ${\cal D}^{(-)}$ are Killing spinors is similar to that for the ${\cal D}^{(+)}$ operator.  In particular, if ${\cal D}^{(-)}\eta_-=0$, then one can show that
 \begin{eqnarray}
{\tilde{\nabla}}^i {\tilde{\nabla}}_i \parallel \eta_-\parallel^2
+ h^i {\tilde{\nabla}}_i \parallel \eta_-\parallel^2+\tilde \nabla^i h_i \parallel \eta_-\parallel^2
= 2 \parallel \nabla^{(-)}\eta_- \parallel^2+  \parallel{\cal{A}}^{(-)}\eta_-\parallel^2~.
\label{lichident1}
\end{eqnarray}
Using $h=d\log \Delta$, this can be rewritten as
\begin{eqnarray}
 {\tilde{\nabla}}^i {\tilde{\nabla}}_i \big(\Delta \parallel \eta_-\parallel^2\big)
- h^i {\tilde{\nabla}}_i \big( \Delta\parallel \eta_-\parallel^2\big)
= 2\Delta \parallel \nabla^{(-)}\eta_- \parallel^2+ \Delta \parallel{\cal{A}}^{(-)}\eta_-\parallel^2~.
\label{lichident2}
\end{eqnarray}
The maximum principle again implies that the only solutions to this equation are those for which $\Delta \parallel \eta_-\parallel=\mathrm{const}$ and $\eta_-$ are Killing spinors.
Again this is always the case if $M^8$ is compact and the fields are smooth. It should be noted that unlike the case of general IIB horizons where this theorem has been proven using a
partial integration formula \cite{iibhor}, here we have presented a different proof based on the maximum principle. The latter has an advantage as it gives
some additional information regarding the length of the Killing spinor $\eta_-$.  Combining the results of this section with those of the previous
one,  we have established that if $M^8$ and the fields satisfy the requirements for the maximum principle to apply, then
\begin{eqnarray}
\nabla^{(\pm)}\eta_\pm=0~,~~~{\cal A}^{(\pm)}\eta_\pm=0\Longleftrightarrow {\cal D}^{(\pm)}\eta_\pm=0~,
\label{ads2lich}
\end{eqnarray}
and that
\begin{eqnarray}
\parallel\eta_+\parallel=\mathrm{const}~,~~~~\Delta\parallel\eta_-\parallel=\mathrm{const}~.
\end{eqnarray}

 \subsection{Counting supersymmetries again}

 The number of supersymmetries of $AdS_2$ backgrounds is
 \begin{eqnarray}
 N=N_-+N_+
 \end{eqnarray}
 where
 \begin{eqnarray}
N_\pm=\mathrm{dim}\,\mathrm{Ker}(\nabla^{(\pm)},  {\cal A}^{(\pm)})~.
\label{nk}
\end{eqnarray}
Using the correspondence between the Killing spinors and zero modes of the ${\cal D}^{(\pm)}$ operators in (\ref{ads2lich}), we conclude that
\begin{eqnarray}
N=\mathrm{dim}\,\mathrm{Ker} {\cal D}^{(-)}+\mathrm{dim}\,\mathrm{Ker} {\cal D}^{(+)}~.
\end{eqnarray}
As for near horizon geometries \cite{iibhor}, one can prove that  $\mathrm{dim}\, \mathrm{Ker}\,{\cal D}^{(+)}{}^\dagger=\mathrm{dim}\, \mathrm{Ker}\,{\cal D}^{(-)}$.
This is done by a direct observation upon comparing the adjoint of ${\cal D}^{(+)}$ with ${\cal D}^{(-)}$.  As a result for $M^8$ compact without boundary,
we find that
\begin{eqnarray}
N=\mathrm{Index} ({\cal D}^{(+)})+ 2 \mathrm{dim}\,\mathrm{Ker} {\cal D}^{(-)}= 2\big(N_-+\mathrm{Index} (D)\big)~,
\end{eqnarray}
where $D$ is the Dirac operator twisted with $\lambda^{{1\over2}}$. The index of ${\cal D}^{(+)}$ is twice the index of $D$ because
they have the same principal symbol and ${\cal D}^{(+)}$ acts on two copies of the Majorana-Weyl representation of $M^8$.  This
establishes both (\ref{susyc}) and (\ref{indexf}) for $AdS_2$ backgrounds.

Furthermore, if $M^8$ is compact without boundary with a $\eta_-$ Killing spinor,  one can explicitly construct a $\eta_+$ Killing spinor
by setting $\eta_+=\Gamma_+\Theta_-\eta_-$.   This is because if $M^8$ is compact without boundary and the fluxes do not vanish,  then $\mathrm{Ker} \Theta_-=\{0\}$.
The proof of this statement is similar to that demonstrated in \cite{iibhor} for near horizon geometries and so it will not be repeated here.

We have shown that the number of supersymmetries preserved by $AdS_2$ backgrounds is even. Apart from this, there are additional restrictions on $N$. In particular,
it has been shown in \cite{n31, n29} that if a IIB background preserves more than 28 supersymmetries, $N>28$, then it is maximally supersymmetric.  Moreover,
the maximally supersymmetric solutions and the   solutions  preserving 28 supersymmetries have been classified in \cite{n32} and \cite{n28}, respectively, and they do not include $AdS_2$
backgrounds. From this, one concludes that $N\leq 26$. One can also adapt the proof of \cite{jose} to this case to demonstrate that all $AdS_2$ backgrounds preserving
more than 16 supersymmetries are homogeneous. This in particular implies that the IIB scalars are constant for all these backgrounds.

\section{$AdS_3 $: Local analysis}

\subsection{Fields, Bianchi identities  and field equations}

The fields restricted on the spatial horizon section ${\cal S}={\mathbb{R}}\times_w M^7$ are
 \begin{eqnarray}
ds^2({\cal S}) &=&  A^2 dz^2 +ds^2(M^7)~,~~~\tilde F^3=A dz\wedge Y~,~~~\tilde F^5= - \star_7 Y~,
\cr
\tilde G^1 &=&  A \Phi dz~,~~~\tilde G^3 = H~,~~~\tilde P^1 = \xi~.
 \end{eqnarray}
Moreover, we have that $h = -{2 \over \ell}dz -2 A^{-1} dA$ and $\Delta=X=L=0$.

Substituting these into the Bianchi identities (\ref{ba0}) and (\ref{ba}), we find that
\begin{align}
 dY &= -3 d\log A \wedge Y + \frac{i}{8} \left( \overline{\Phi} H - \Phi \overline{H} \right)~,
\nonumber \\
  d \Phi &= 3 \Phi d\log A + i \Phi Q - \overline{\Phi} \xi~,
\nonumber  \\
 d *_7 Y &= -\frac{i}{8} H \wedge \overline{H}~,
\nonumber \\
 dH &= i Q \wedge H - \xi \wedge \overline{H}~,
\nonumber \\
 d\xi &= 2 i Q \wedge \xi~,
\nonumber \\
 dQ &= -i \xi \wedge \overline{\xi}~.
\end{align}
In addition the field equations (\ref{feq1})-(\ref{feq7}) give
\begin{align}
 \nabla^i H_{i j k} &= -3 A^{-1} \partial^i A H_{i j k} + i Q^i G_{i j k} + P^i H_{i j k} + 4 i \Phi Y_{j k} + \frac{i}{3} \epsilon_{j k i_1 i_2 \ell_1 \ell_2 \ell_3} Y^{i_1 i_2} H^{\ell_1 \ell_2 \ell_3}~,
 \nonumber \\
 \nabla^i \xi_i &= -3 A^{-1} \partial^i A \xi_i + 2 i Q^i \xi_i - \frac{1}{24} H^2 - \frac{1}{4} \Phi^2~,
 \nonumber \\
 A^{-1} \nabla^2 A &= 2 Y^2 + \frac{3}{8} \parallel \Phi \parallel ^2 + \frac{1}{48} \parallel H \parallel^2 - 2\ell^{-2}A^{-2} - 2 ( d\log A ) ^2~,
 \label{ads3feq1}
 \nonumber \\
  R^{(7)}_{i j} &= 3 A^{-1} \nabla_{i} \nabla_{j} A + 2 Y^2 \delta_{i j} - 8 Y^2_{i j}
 \nonumber \\
 & \qquad \qquad + \frac{1}{4} H_{( i}{}^{k \ell} \overline{H}_{j ) k \ell} + \frac{1}{8} \parallel \Phi \parallel^2 \delta_{i j} - \frac{1}{48} \parallel H \parallel^2 \delta_{i j} + 2 \xi_{(i}
 \bar\xi_{j)} ,
\end{align}
where $R^{(7)}$ is the Ricci tensor of $M^7$. From now on, $\nabla$ will denote the Levi-Civita covariant derivative on $M^{10-n}$.
The Ricci scalar of $M^7$ is given by
\begin{align}
 R^{(7)} &= 3 A^{-1} \nabla^2 A + 6 Y^2 + \frac{5}{48}\parallel H \parallel^2 + \frac{7}{8} \parallel \Phi \parallel ^2 + 2 \left| \xi \right| ^2
 \nonumber \\
 &= -\frac{6}{\ell^2} A^{-2}- 6 (A^{-1}dA) ^2 + 12 Y^2 + 2 \parallel \Phi \parallel ^2 + \frac{1}{6} \parallel H \parallel^2 + 2 \parallel \xi \parallel^2 .
\end{align}

\subsection{The warp factor is no-where vanishing}

One of the consequences of the field equations is that the warp factor $A$ is no-where vanishing.  One can show that this follows from the field equation
(\ref{ads3feq1}) using an argument similar to that presented for the $AdS_2$ backgrounds.

\subsection{Solution of Killing spinor equations}

To integrate the KSEs along the $AdS_3$ directions, it suffices to integrate the horizon KSEs (\ref{horkses}) along the $z$ coordinate. For this consider first the
gravitino KSE. Evaluating the expression along the $z$-coordinate, we find
\begin{eqnarray}
\partial_z \eta_\pm = \Xi_\pm \eta_\pm~,
\label{zads3}
\end{eqnarray}
where
\begin{eqnarray}
\Xi_\pm=\mp{1\over 2\ell}-{1\over 2} \Gamma_z \slashed{\partial}A  \pm {i\over 4} A \slashed{Y} +\left({1\over 96} A \Gamma_z \slashed{H} \pm {3\over16}A \Phi\right) C*~.
\end{eqnarray}
Observe that\footnote{The gamma matrices labeled by $AdS_n$ coordinates, like $\Gamma_z$,  are in a frame basis.}
\begin{eqnarray}
\Xi_+= A \Gamma_z \Theta_+~,~~~\Xi_-={1\over \ell}+ A \Gamma_z \Theta_-~.
\end{eqnarray}
Next differentiating (\ref{zads3}) and comparing the resulting expression with the integrability conditions (\ref{alg1}), one finds that
\begin{eqnarray}
\partial^2_z \eta_\pm\pm {1\over \ell} \partial_z\eta_\pm=0~,
\end{eqnarray}
which can be solved to give
\begin{eqnarray}
\eta_\pm=\sigma_\pm+ e^{\mp {z\over\ell}} \tau_\pm~,
\label{ads3eta}
\end{eqnarray}
where
\begin{eqnarray}
\Xi_\pm \sigma_\pm=0~,~~~\Xi_\pm \tau_\pm=\mp{1\over\ell} \tau_\pm~,
\end{eqnarray}
with both $\sigma_\pm$ and $\tau_\pm$ $z$-independent spinors. The latter conditions are additional independent algebraic KSEs.

Although we have solved along the $z$ direction, there are potentially additional conditions that can arise from
mixed integrability conditions along the $z$-direction and the remaining directions in ${\cal S}$.  However, it can be shown after
some computation that this is not the case. Furthermore, the dilatino KSEs in (\ref{horkses}) restrict on the $\sigma_\pm$ and $\tau_\pm$ spinors
in a straightforward manner.  This completes the integration of the KSEs along all $AdS_3$ directions. The remaining independent KSEs, which are localized
on $M^7$, are
\begin{eqnarray}
&&\nabla^{(\pm)}_i \sigma_\pm=0~,~~~\nabla^{(\pm)}_i\tau_\pm=0~,
\cr
&&{\cal A}^{(\pm)} \sigma_\pm=0~,~~~{\cal A}^{(\pm)} \tau_\pm=0~,
\cr
&&{\cal B}^{(\pm)} \sigma_\pm=0~,~~~{\cal C}^{(\pm)} \tau_\pm=0~,
\label{ads3indkse}
\end{eqnarray}
where
\begin{eqnarray}
&&\nabla_i^{(\pm)}=\nabla_i+\Psi_i^{(\pm)}~,~~~{\cal A}^{(\pm)}=\mp{1\over4} \Phi \Gamma_z+{1\over24} \slashed {H}+ \slashed {\xi} C*~,
\cr
&&{\cal B}^{(\pm)}=\Xi_\pm~,~~~{\cal C}^{(\pm)}=\Xi_\pm\pm {1\over\ell}~,
\label{ppalgads3}
\end{eqnarray}
and
\begin{align}
 \Psi^{( \pm )}_{i} &= \pm \frac{1}{2 } \partial_{i} \log A - \frac{i}{2} Q_{i} \pm \frac{i}{4} \left( \slashed{\Gamma Y} \right) _i \Gamma_z \mp \frac{i}{2} \slashed{Y}_i \Gamma^z
\nonumber \\
 & \qquad \qquad + \left( -\frac{1}{96} \left( \slashed{\Gamma H} \right) _i + \frac{3}{32} \slashed{H}_i \mp \frac{1}{16} \Phi \Gamma_{z i} \right) C * .
\end{align}
Therefore, there are four sets of three independent KSEs on $M^7$. Having found a solution to the above equations, one can substitute in (\ref{ads3eta}) and then
in (\ref{st1}) to find the Killing spinors for the $AdS_3\times_w M^7$ background.

\subsection{Counting supersymmetries}

It is straightforward to observe that if one has an either a $\sigma_-$ or a $\tau_-$ solution, then
\begin{eqnarray}
\sigma_+=A^{-1} \Gamma_z\Gamma_+\sigma_-~,~~~\tau_+=A^{-1} \Gamma_z\Gamma_+\tau_-~,
\end{eqnarray}
are also solutions of the independent KSEs (\ref{ads3indkse}).  Conversely, if either $\sigma_+$ or $\tau_+$ are solutions,
then
\begin{eqnarray}
\sigma_-=A \Gamma_z\Gamma_-\sigma_+~,~~~\tau_-=A \Gamma_z\Gamma_-\tau_+~,
\end{eqnarray}
are also solutions to the KSEs (\ref{ads3indkse}).  Therefore, we have that the number of Killing spinors $N$ of the $AdS_3$  backgrounds are
\begin{eqnarray}
N&=&2 \big(\mathrm{dim}\, \mathrm{Ker}(\nabla^{(-)}, {\cal A}^{(-)}, {\cal B}^{(-)})+ \mathrm{dim}\, \mathrm{Ker}(\nabla^{(-)}, {\cal A}^{(-)}, {\cal C}^{(-)})\big)
\cr
&=&2\big(\mathrm{dim}\, \mathrm{Ker}(\nabla^{(+)}, {\cal A}^{(+)},  {\cal B}^{(+)})+ \mathrm{dim}\, \mathrm{Ker}(\nabla^{(+)}, {\cal A}^{(+)}, {\cal C}^{(+)})\big)~.
\label{nads3}
\end{eqnarray}
Thus the $AdS_3$ backgrounds preserve an even number of supersymmetries.  This proves the formula for $N$ in (\ref{susyc}) for $AdS_3$ backgrounds.

The number of supersymmetries $N$ of $AdS_3$ backgrounds are further restricted. It follows from the results of \cite{n31, n29, n28} that there are no
supersymmetric $AdS_3$ backgrounds preserving more than 28 supersymmetries.  As a result, $N\leq 26$.

\section{$AdS_3$: Global analysis}

The main task here is to show the formula (\ref{indexf}) for counting the number of supersymmetries of $AdS_3$ backgrounds.  For this,
we have to show that there is a 1-1 correspondence between Killing spinors and zero modes of a Dirac-like operator on $M^7$.

\subsection{A Lichnerowicz type theorem for $\tau_+$ and $\sigma_+$}

To prove that the zero modes of a Dirac-like operator on $M^7$ are Killing spinors,
one has to determine an appropriate Dirac-like operator on $M^7$. The naive Dirac-like operator which
one can construct from contracting $\nabla^{(\pm)}$ with a gamma matrix is not suitable.  Instead, let us modify
the parallel transport operators of the gravitino KSE as
\begin{eqnarray}
&&\check \nabla_i^{(+)}=\nabla_i^{(+)}+ q \Gamma_{zi} A^{-1} {\cal B}^{(+)}~,
\cr
&&\hat \nabla_i^{(+)}=\nabla_i^{(+)}+ q \Gamma_{zi} A^{-1} {\cal C}^{(+)}~,
\label{ads3ch}
\end{eqnarray}
on $\sigma_+$ and $\tau_+$, respectively, where $q$ is a number which later will be set to $1/7$. It is clear that if either
$\sigma_+$ or $\tau_+$ are Killing spinors, they are also parallel with respect to the above covariant derivatives.

Since the analysis that follows is similar for $\sigma_+$ and $\tau_+$, it is convenient to present it in a unified way.
For this write both (\ref{ads3ch}) as
\begin{eqnarray}
\mathbb{D}_i^{(+)}= \nabla^{(+)}_i+ q \Gamma_{zi} A^{-1} \mathbb{B}^{(+)}~,
\end{eqnarray}
where
\begin{eqnarray}
\mathbb{B}^{(+)}=-{c\over 2\ell}-{1\over 2} \Gamma_z \slashed{\partial}A  + {i\over 4} A \slashed{Y}+\left({1\over 96} A \Gamma_z \slashed{H} + {3\over16}A \Phi\right) C*~,
\end{eqnarray}
and $c=1$ when acting on $\sigma_+$ and $c=-1$ when acting on $\tau_+$, ie either $\mathbb{B}^{(+)}={\cal B}^{(+)}$ or $\mathbb{B}^{(+)}={\cal C}^{(+)}$, respectively.

Next define the modified Dirac-like operators
\begin{eqnarray}
\mathscr{D}^{(+)}\equiv \Gamma^i \mathbb{D}_i^{(+)}=\Gamma^i \nabla_i+\Sigma^{(+)}~,
\end{eqnarray}
where
\begin{eqnarray}
\Sigma^{(+)}&=& \frac{7 q c}{2 \ell} A^{-1}\Gamma_z + \frac{1 + 7 q}{4 } \slashed{\partial} \log A^2 - \frac{i}{2} \slashed{Q} + \frac{3 i - 7 i q}{4} \slashed{Y} \Gamma_z
\cr
&&+ \big(\frac{5 - 7 q}{96} \slashed{H} +{7-21 q\over 16} \Phi\big)C *~.
\end{eqnarray}
It turns that $\mathscr{D}^{(+)}$ is suitable to formulate a maximum principle on the length square of $\sigma_+$ and $\tau_+$.  In particular, suppose that
$\chi_+$ is a zero mode for $\mathscr{D}^{(+)}$, ie $\mathscr{D}^{(+)}\chi_+=0$, where $\chi_+=\sigma_+$ for $c=1$ while $\chi_+=\tau_+$ for $c=-1$. Then
after some Clifford algebra, that is presented in appendix \ref{clifads3}, which requires the use of field equations and for $q=1/7$, one can establish the identity
\begin{eqnarray}
 {\nabla}^2 \left\| \chi_+ \right\| ^2 &+& 3 A^{-1} \partial^{i} A \partial_{i} \left\| \chi_+ \right\| ^2 = 2 \left\| \mathbb{D}^{( + )} \chi_+ \right\| ^2
 \cr
 &&+ \frac{16}{7} A^{-2} \left\| \mathbb{B}^{(+)} \chi_+ \right\| ^2 + \left\| \cal{A}^{(+)} \chi_+ \right\| ^2~.
 \label{ads3maxp}
\end{eqnarray}
Assuming that $M^7$ satisfies the requirements of the Hopf maximum principle, eg for $M^7$  compact and smooth fields,  the above equation
implies that $\chi_+$ is a Killing spinor and that the length $\parallel\chi_+\parallel=\mathrm{const}$.

To summarize, we have shown that
\begin{eqnarray}
\nabla^{(+)}_i\sigma_+=0~,~~{\cal B}^{(+)}\sigma_+=0~,~~{\cal A}^{(+)}\sigma_+=0 \Longleftrightarrow \mathscr{D}^{(+)}\sigma_+=0~;~c=1~,
\cr
\nabla^{(+)}_i\tau_+=0~,~~{\cal C}^{(+)}\tau_+=0~,~~{\cal A}^{(+)}\tau_+=0 \Longleftrightarrow \mathscr{D}^{(+)}\tau_+=0~;~c=-1~,
\end{eqnarray}
and that
\begin{eqnarray}
\parallel\sigma_+\parallel=\mathrm{const}~,~~~\parallel\tau_+\parallel=\mathrm{const}~.
\end{eqnarray}

\subsection{A Lichnerowicz type theorem for $\tau_-$ and $\sigma_-$}

A similar theorem to that presented in the previous section can be established for $\tau_-$ and $\sigma_-$ spinors. One can define
the operators $\mathbb D^{(-)}$ and  $\mathscr{D}^{(-)}$  and repeat the analysis.  Alternatively, one can observe that
if $\chi_+$ is a zero mode of the $\mathscr{D}^{(+)}$ operator, then $\chi_-=A \Gamma_{z-} \chi_+$ is a zero mode
of the $\mathscr{D}^{(-)}$ operator, where $\chi_-$ is either $\sigma_-$ or $\tau_-$.  Since the same relation holds
between $\chi_+$ and $\chi_-$ Killing spinors, one can establish a maximum principle for $\chi_-$ spinor.  The formula
is that given in (\ref{ads3maxp}) after setting  $\chi_+= A^{-1} \Gamma_{z+} \chi_-$. Therefore provided the requirements
of Hopf maximum principle are satisfied, one establishes
\begin{eqnarray}
\nabla^{(-)}_i\sigma_-=0~,~~{\cal B}^{(-)}\sigma_-=0~,~~{\cal A}^{(-)}\sigma_-=0 \Longleftrightarrow \mathscr{D}^{(-)}\sigma_-=0~;~c=1~,
\cr
\nabla^{(-)}_i\tau_-=0~,~~{\cal C}^{(-)}\tau_-=0~,~~{\cal A}^{(-)}\tau_-=0 \Longleftrightarrow \mathscr{D}^{(-)}\tau_-=0~;~c=-1~,
\end{eqnarray}
and that
\begin{eqnarray}
A^{-2}\parallel\sigma_-\parallel^2=  \mathrm{const}~,~~~A^{-2} \parallel\tau_-\parallel^2=\mathrm{const}~,
\end{eqnarray}
where
\begin{eqnarray}
\mathscr{D}^{(-)}=\Gamma^i \nabla_i+\Sigma^{(-)}~,
\end{eqnarray}
and
\begin{eqnarray}
\Sigma^{(-)}&=& -\frac{7 q c}{2\ell  } A^{-1}\Gamma_z + \frac{-1 + 7 q}{4 } \slashed{\partial} \log A^2 - \frac{i}{2} \slashed{Q} - \frac{3 i - 7 i q}{4} \slashed{Y} \Gamma_z
\cr
&&+ \big(\frac{5 - 7 q}{96} \slashed{H} -{7-21 q\over 16} \Phi\big)C *~.
\label{ads3maxpm}
\end{eqnarray}

\subsection{Counting supersymmetries again}

The proof of the relation between Killing spinors and the zero modes of the Dirac-like operators $\mathscr{D}^{(\pm)}$ allows us
to re-express the number of supersymmetries $N$ in (\ref{nads3}) preserved by $AdS_3$ backgrounds as
\begin{eqnarray}
N&=&2 \big(\mathrm{dim}\, \mathrm{Ker}\,  \mathscr{D}^{(-)}_{c=1} +\mathrm{dim}\, \mathrm{Ker} \,\mathscr{D}^{(-)}_{c=-1}\big)
\cr
&=&2 \big(\mathrm{dim}\, \mathrm{Ker} \,\mathscr{D}^{(+)}_{c=1}+\mathrm{dim}\, \mathrm{Ker}\,  \mathscr{D}^{(+)}_{c=-1}\big)~,
\end{eqnarray}
which establishes (\ref{indexf}) for $AdS_3$.

\section{$AdS_4$: Local analysis}

\subsection{Fields, Bianchi identities and field equations}

The field on ${\cal S}$ are
\begin{eqnarray}
&&ds^2({\cal S})=A^2 (dz^2+ e^{2z/\ell} dx^2)+ds^2(M^6)~,~~~\tilde F^3=A^2 e^{z/\ell} dz\wedge dx\wedge Y~,~~~
\cr
&&\tilde F^5=\star_6 Y~,~~~\tilde G^3=H~,~~~P=\xi~,
\end{eqnarray}
with $h = -{2 \over \ell}dz -2 A^{-1} dA$ and $\Delta=X=L=0$.
Substituting these into the Bianchi and field equations on ${\cal S}$ in section \ref{horbf},
the conditions reduce on $M^6$ as follows. The Bianchi identities give

\begin{align}
 d(A^4Y) &= 0~,~~~
 \tilde{\nabla}^{i} Y_{i} = -\frac{i}{288} \epsilon^{i_1 i_2 i_3 j_1 j_2 j_3} H_{{i_1} {i_2} {i_3}} \overline{H}_{{j_1} {j_2} {j_3}}~,
\nonumber \\
 dH &= i Q \wedge H - \xi \wedge \overline{H}~,
\nonumber \\
 d\xi &= 2 i Q \wedge \xi~,
\nonumber \\
 dQ &= -i \xi \wedge \overline{\xi}~.
\end{align}
Therefore the Bianchi identities imply that $A^4 Y$ is a closed 1-form and that $H\wedge \overline H$ represents a trivial cohomology class in $M^6$.

The Einstein equation on ${\cal S}$ gives
\begin{equation}
  A^{-1}{\nabla}^2 A = 4 Y^2 + \frac{1}{48} \parallel H \parallel^2 - \frac{3}{\ell^2 A^2} - 3 (A^{-1} dA )^2,
  \label{ads4feq1}
\end{equation}
and
\begin{align}
 &R^{(6)}_{i j} - 4 A^{-1} {\nabla}_{i} {\nabla}_{j} A - 4 Y^2 \delta_{i j} + 8  Y_{i} Y_{j}
 \\ \nonumber
 & \qquad \qquad - \frac{1}{4} H_{( i}{}^{k \ell} \overline{H}_{j ) k \ell} + \frac{1}{48} \parallel H \parallel^2 \delta_{i j} - 2 \xi_{( i} \overline{\xi}_{j )} = 0~,
\end{align}
where $R^{(6)}$ is the Ricci tensor of $M^6$.  The remaining field equations are
\begin{eqnarray}
 \nabla^{i} H_{i j k} &=& -3  \partial^i\log A\, H_{ijk}+i Q^{i} H_{i j k} + \xi^{i} \overline{H}_{i j k}~,~~~
 \cr
 \nabla^{i} \xi_{i} &=&-3  \partial^i\log A\, \xi_i+ 2 i Q^{i} \xi_{i} - \frac{1}{24} H^2 .
\end{eqnarray}
This concludes the reduction of the Bianchi identities and field equations on $M^6$.

\subsubsection{The warp factor is no-where vanishing}
 One consequence of the field equations and in particular of (\ref{ads4feq1}) is that the warp factor $A$ is no-where vanishing.
 The investigation for this is similar to that we have presented for $AdS_3$ and so we shall not repeat the argument here.

\subsection{Solution of KSEs}

The integration of the KSEs along the $z$-coordinate proceeds as in the $AdS_3$. In particular repeating the argument as in the
$AdS_3$ case, one finds that
\begin{eqnarray}
\eta_\pm=\phi_\pm+ e^{\mp z/\ell} \chi_\pm~,
\end{eqnarray}
where
\begin{eqnarray}
\Xi_\pm \phi_\pm=0~,~~~\Xi_\pm \chi_\pm=\mp{1\over\ell} \chi_\pm~,~~~{\cal A}^{(\pm)}\phi_\pm=0~,~~~{\cal A}^{(\pm)}\chi_\pm=0~,
\end{eqnarray}
and
\begin{eqnarray}
&&\Xi_\pm=\mp{1\over 2\ell}-{1\over 2} \Gamma_z \slashed{\partial} A \pm {i\over 2} A \Gamma_x \slashed{Y} +{1\over 96} A \Gamma_z\slashed{H} C*~,
\cr
&& {\cal A}^{(\pm)}={1\over24} \slashed {H}+ \slashed {\xi} C*~.
\end{eqnarray}
Observe that although ${\cal A}^{(+)}={\cal A}^{(-)}$ as operators, they act on different spaces and so we shall retain the distinct labeling.

Next we integrate the gravitino KSE along the $x$ AdS coordinate to obtain
\begin{eqnarray}
\eta_+=\sigma_+-{1\over \ell} x \Gamma_x \Gamma_z \tau_++ e^{-z/\ell} \tau_+~,~~~\eta_-=\sigma_-+ e^{z/\ell} (-{1\over \ell} x \Gamma_x \Gamma_z\sigma_-+\tau_-)~,
\label{ads4eta}
\end{eqnarray}
where
\begin{eqnarray}
\Xi_\pm\sigma_\pm=0~,~~~\Xi_\pm\tau_\pm=\mp {1\over\ell} \tau_\pm~,
\end{eqnarray}
and $\sigma_\pm$ and $\tau_\pm$ depend only on the coordinates of $M^6$.  This completes the integration of the gravitino KSE along all $AdS_4$ directions.  The dilatino
KSE simply restricts on the spinors $\sigma_\pm$ and $\tau_\pm$.  There are no additional conditions arising from integrability conditions
between $AdS_4$ and $M^6$ directions.

Therefore, the remaining independent KSEs on $M^6$ are
\begin{eqnarray}
&&\nabla_i^{(\pm)}\sigma_\pm=0~,~~~\nabla_i^{(\pm)}\tau_\pm=0~,
\cr
&&{\cal A}^{(\pm)} \sigma_\pm=0~,~~~{\cal A}^{(\pm)} \tau_\pm=0~,
\cr
&&{\cal B}^{(\pm)}\sigma_\pm=0~,~~~{\cal C}^{(\pm)}\tau_\pm=0~,
\label{ads4kses}
\end{eqnarray}
where
\begin{eqnarray}
\nabla_i^{(\pm)}=\nabla_i+ \Psi^{(\pm)}_i~,~~~{\cal B}^{(\pm)}=\Xi_\pm~,~~~{\cal C}^{(\pm)}=\Xi_\pm \pm {1\over\ell}~,
\label{ppalgads4}
\end{eqnarray}
and
\begin{eqnarray}
\Psi^{( \pm )}_{i} &=& \pm \frac{1}{2 } \partial_{i} \log A - \frac{i}{2} Q_{i} \mp \frac{i}{2} \left( \slashed{\Gamma Y} \right) _i \Gamma_{xz} \pm \frac{i}{2} Y_i \Gamma_{xz}
 \cr
 &&~~~+ \left( -\frac{1}{96} \left( \slashed{\Gamma H} \right) _i + \frac{3}{32} \slashed{H}_i  \right) C * .
\end{eqnarray}
This concludes the solution of the KSEs on $AdS_4$ and their reduction on $M^6$.

\subsection{Counting supersymmetries}

As for  $AdS_3$ backgrounds there are Clifford algebra operators which intertwine between the different KSEs on $M^6$.
In particular observe that  if $\sigma_\pm$ is a solution to the KSEs, then
\begin{eqnarray}
\tau_\pm=\Gamma_z \Gamma_x \sigma_\pm
\end{eqnarray}
is also a solution, and vice versa.   Furthermore as for $AdS_3$,   if either $\sigma_-$ or $\tau_-$ is a solution, so is
\begin{eqnarray}
\sigma_+=A^{-1} \Gamma_+ \Gamma_z \sigma_-~,~~~\tau_+=A^{-1} \Gamma_+ \Gamma_z \tau_-~.~~~
\label{pmrelads4}
\end{eqnarray}
Similarly, if either  $\sigma_+$ or  $\tau_+$ is a solution, so is
\begin{eqnarray}
\sigma_-=A \Gamma_- \Gamma_z \sigma_+~,~~~\tau_-=A \Gamma_- \Gamma_z \tau_+~.~~~
\label{mprelads4}
\end{eqnarray}
From the above relations one concludes that the $AdS_4\times_w M^{6}$ backgrounds preserve
\begin{eqnarray}
N= 4 \,\,\mathrm{dim}\, \mathrm {Ker}(\nabla^{(\pm)}, {\cal A}^{(\pm)}, {\cal B}^{(\pm)})= 4 \,\,\mathrm{dim}\, \mathrm {Ker}(\nabla^{(\pm)}, {\cal A}^{(\pm)}, {\cal C}^{(\pm)})~,
\label{nads4}
\end{eqnarray}
for either $+$  or $-$  choice of sign. This confirms (\ref{susyc}) for the $AdS_4$ backgrounds.

The number of supersymmetries $N$ of $AdS_4$ backgrounds are further restricted. It is a consequence of \cite{n32, n31, n29, n28} that there are no $AdS_4$
backgrounds with $N\geq 28$ supersymmetries.  Therefore $N\leq 24$.

\section{$AdS_4$:  Global analysis}

\subsection{A Lichnerowicz type theorem for $\tau_{\pm}$ and $\sigma_{\pm}$}

To prove the formula (\ref{indexf}) for $AdS_4$ backgrounds, we have to demonstrate a Lichnerowicz type theorem which states that there is a 1-1 correspondence between
Killing spinors and the zero modes of Dirac-like operators on $M^6$ coupled to fluxes.  The proof is similar to that we have presented for the $AdS_3$ backgrounds.
However, the operators involved in the $AdS_4$ case are different and so the proof is not a mere repetition.

We shall present the proof of the Lichnerowicz type theorem for $\sigma_+$ and $\tau_+$ spinors. The proof for the other pair $\sigma_-$ and $\tau_-$ follows
as a consequence.  It is also convenient to do the computations simultaneously for both $\sigma_+$ and $\tau_+$ spinors which from now on we shall call collectively $\chi_+$.

To begin let us define the operator
\begin{eqnarray}
\mathbb{D}_i^{(+)}= \nabla^{(+)}_i+ q \Gamma_{zi} A^{-1} \mathbb{B}^{(+)}
\end{eqnarray}
where
\begin{eqnarray}
\mathbb{B}^{(+)}=-{c\over 2\ell}-{1\over 2} \Gamma_z \slashed{\partial} A  - {i\over 2} A \slashed{Y} \Gamma_x+{1\over 96} A \Gamma_z \slashed{H} C*
\end{eqnarray}
and $c=1$ when acting on $\sigma_+$ and $c=-1$ when acting on $\tau_+$, ie either $\mathbb{B}^{(+)}={\cal B}^{(+)}$ or $\mathbb{B}^{(+)}={\cal C}^{(+)}$, respectively.
  It is clear
from this that if $\chi_+$ is a Killing spinor, then it is parallel with respect to $\mathbb{D}^{(+)}$.

Next define the modified Dirac-like operator
\begin{eqnarray}
\mathscr{D}^{(+)}\equiv \Gamma^i \mathbb{D}_i^{(+)}=\Gamma^i \nabla_i+\Sigma^{(+)}~,
\end{eqnarray}
where
\begin{eqnarray}
\Sigma^{(+)}= \frac{3 q c}{2 } A^{-1}\Gamma_z + \frac{1 + 6 q}{4 } \slashed{\partial} \log A^2 - \frac{i}{2} \slashed{Q} + (2 i - 3 i q) \slashed{Y} \Gamma_{zx}  + \frac{1 -  q}{16} \slashed{H} C *~.
\end{eqnarray}
Next suppose that $\chi_+$ is a zero mode of  $\mathscr{D}^{(+)}$, ie $\mathscr{D}^{(+)}\chi_+=0$. Then after some Clifford algebra
 computation, which has been presented in appendix \ref{clifads4}, $q=1/3$,  and the use of field equations, one can establish the identity
\begin{eqnarray}
 {\nabla}^2 \left\| \chi_+ \right\| ^2 &+& 4 A^{-1} \partial^{i} A \partial_{i} \left\| \chi_+ \right\| ^2 = 2 \left\| \mathbb{D}^{( + )} \chi_+ \right\| ^2
 \cr
 &&+ \frac{16}{3} A^{-2} \left\| \mathbb{B}^{(+)} \chi_+ \right\| ^2 + \left\| \cal{A}^{(+)} \chi_+ \right\| ^2~.
 \label{ads4maxpp}
\end{eqnarray}
Assuming that the  requirements of the Hopf maximum principle are satisfied, eg for $M^6$  compact and smooth fields,  the above equation
implies that $\chi_+$ is a Killing spinor and that the length $\parallel\chi_+\parallel=\mathrm{const}$.

A similar formula to (\ref{ads4maxpp}) can be established for $\sigma_-$ and $\tau_-$ spinors. However, it is not necessary to do an independent computation.  We have seen that
if $\sigma_+$ and $\tau_+$ solve the KSEs, then $\sigma_-=A \Gamma_{-z} \sigma_+$ and  $\tau_-=A \Gamma_{-z} \tau_+$ also solve the KSEs. Similarly
if $\chi_+$ is a zero mode of $\mathscr{D}^{(+)}$, then $\chi_-=A \Gamma_{-z} \chi_+$ is a zero mode of $\mathscr{D}^{(-)}$, where
\begin{eqnarray}
\mathscr{D}^{(-)}=\Gamma^i \nabla_i+ \Sigma^{(-)}~,
\end{eqnarray}
and
\begin{eqnarray}
\Sigma^{(-)}=-\frac{3q c}{2 \ell} A^{-1}\Gamma_z + \frac{-1+6q}{4 } \slashed{\partial} \log A^2 - \frac{i}{2} \slashed{Q} - (2 i - 3 i q) \slashed{Y} \Gamma_{zx}  + \frac{1 -  q}{16} \slashed{H} C *~.
\end{eqnarray}
To summarize, we have shown that
\begin{eqnarray}
\nabla^{(\pm)}_i\sigma_\pm=0~,~~{\cal B}^{(\pm)}\sigma_\pm=0~,~~{\cal A}^{(\pm)}\sigma_\pm=0 \Longleftrightarrow \mathscr{D}^{(\pm)}\sigma_\pm=0~;~c=1~,
\cr
\nabla^{(+)}_i\tau_\pm=0~,~~{\cal C}^{(\pm)}\tau_\pm=0~,~~{\cal A}^{(\pm)}\tau_\pm=0 \Longleftrightarrow \mathscr{D}^{(\pm)}\tau_\pm=0~;~c=-1~,
\end{eqnarray}
and that
\begin{eqnarray}
\parallel\sigma_+\parallel&=&\mathrm{const}~,~~~\parallel\tau_+\parallel=\mathrm{const}~,
\cr
A^{-2}\, \parallel\sigma_-\parallel^2&=&  \mathrm{const}~,~~~A^{-2}\,\parallel\tau_-\parallel^2= \mathrm{const}~.
\end{eqnarray}
This concludes the proof of the 1-1 correspondence between Killing spinors and zero modes of Dirac-like operators on $M^6$.

\subsection{Counting supersymmetries again}

We are ready now to establish (\ref{indexf}) for $AdS_4$ backgrounds. Provided that the data satisfy the requirements
of Hopf maximum principle, we have that
\begin{eqnarray}
N= 4 \,\,\mathrm{dim}\, \mathrm {Ker}(\nabla^{(-)}, {\cal A}^{(-)}, {\cal B}^{(-)})=4\,\, \mathrm{dim}\, \mathrm {Ker} \mathscr{D}^{(-)}_{c=1}~,
\end{eqnarray}
which applies to $\sigma_-$ spinors which confirms (\ref{indexf}).  A similar formula is valid for the three other choices of spinors.

\section{$AdS_5$: Local analysis}

\subsection{Fields, Bianchi identities and field equations}
The fields on the horizon section ${\cal S}$ are
\begin{eqnarray}
ds^2({\cal S})&=& A^2( dz^2+e^{2z\over\ell} (dx^2+dy^2))+ ds^2(M^5)~,~~~G=H~,~~P=\xi~,
\cr
\tilde F^3&=& e^{2z\over\ell} A^3 dz\wedge dx\wedge dy\, Y~,~~~\tilde F^5=-d\mathrm{vol}(M^5) Y~,
\end{eqnarray}
and $h=-{2\over \ell} dz-2 d\log A$ and $\Delta=X=L=0$.

Substituting the above fields into the Bianchi identities (\ref{ba0}) and (\ref{ba}), we find
\begin{eqnarray}
 d(A^5Y )&=&0~,~~~dH = i Q \wedge H - \xi \wedge \overline{H}~,
 \cr
 d\xi &=& 2 i Q \wedge \xi~,~~~ dQ = -i \xi \wedge \overline{\xi} .
\end{eqnarray}
Clearly, $Y$ is proportional to $A^{-5}$. Similarly, the field equations (\ref{feq1})-(\ref{feq7}) give
\begin{eqnarray}
 {\nabla}^{i} H_{i j k} &=& -5 \partial^i \log A \, H_{i j k} + i Q^{i} H_{i j k} + \xi^{i} \overline{H}_{i j k}~,
 \cr
 {\nabla}^{i} \xi_{i} &=& -5 \partial^i \log A \, \xi_i + 2 i Q^{i} \xi_{i} - \frac{1}{24} H^2~,
\cr
 A^{-1} {\nabla}^2 A &=& 4 Y^2 + \frac{1}{48} \parallel H \parallel^2 - \frac{4}{\ell^2} A^{-2} - 4 (d\log A)^2 ,
\cr
  {R}^{(5)}_{i j}&=&  5 A^{-1} {\nabla}_{i} {\nabla}_{j} A + 4 Y^2 \delta_{i j}
  \cr
 &+& \frac{1}{4} H_{( i}{}^{k \ell} \overline{H}_{j ) k \ell} - \frac{1}{48} \parallel H \parallel^2 \delta_{i j} + 2 \xi_{( i} \overline{\xi}_{j )}~.
 \label{ads5feq}
\end{eqnarray}
This concludes the analysis of Bianchi and field equations.

\subsubsection{The warp factor is nowhere vanishing}

As in the previous AdS backgrounds, one can show that the warp factor $A$ is no-where vanishing. The argument is based on the third field equation in (\ref{ads5feq}).
\subsection{Solution of KSEs}

Substituting the fields of the previous section into the KSEs of the spatial horizon section  (\ref{horkses}) and after a computation similar to that
described for  $AdS_4$ backgrounds, we find that the Killing spinors can be  expressed as
\begin{eqnarray}
\eta_+=\sigma_+-{1\over\ell} (x \Gamma_x+y \Gamma_y) \Gamma_z \tau_++ e^{-{z\over \ell}} \tau_+~,~~\eta_-=\sigma_-+ e^{{z\over\ell}} \big(-{1\over\ell}
(x \Gamma_x+y \Gamma_y) \Gamma_z \sigma_-+\tau_-\big)~,
\end{eqnarray}
where $\sigma_\pm$ and $\tau_\pm$ depend only on the coordinates of $M^5$. The remaining independent KSEs are
\begin{eqnarray}
&&\nabla_i^{(\pm)}\sigma_\pm=0~,~~~\nabla_i^{(\pm)}\tau_\pm=0~,~~~{\cal A}^{(\pm)}\sigma_\pm=0~,~~~{\cal A}^{(\pm)}\tau_\pm=0~,
\cr
&&{\cal B}^\pm\sigma_\pm=0~,~~~{\cal C}^\pm\tau_\pm=0~,
\label{ads5x}
\end{eqnarray}
where
\begin{eqnarray}
&&\nabla_i^{(\pm)}=  \nabla_i+ \Psi^{(\pm)}_i~,~~~{\cal A}^{(\pm)}={1\over24} \slashed H+\slashed\xi C*~,
\cr
&&{\cal B}^{(\pm)}=\Xi_\pm~,~~~{\cal C}^{(\pm)}=\Xi_\pm\pm {1\over\ell}~,
\end{eqnarray}
and
\begin{eqnarray}
\Psi_i^{(\pm)}&=&\pm{1\over2} \partial_i\log A-{i\over2} Q_i\pm {i\over2} \Gamma_i Y \Gamma_{xyz}+ \big(-{1\over96} (\Gamma\slashed H)_i+{3\over32} \slashed H_i\big) C*
\cr
\Xi_\pm&=&\mp {1\over 2\ell}-{1\over2} \Gamma_z \slashed{\partial} A \pm {i\over2} AY \Gamma_{xy}+ {1\over 96} A \Gamma_z \slashed H C*~.
\end{eqnarray}
This concludes the solution of the KSEs along the $AdS_5$ directions and the identification of remaining independent KSEs.

\subsection{Counting supersymmetries}

To count the number of supersymmetries preserved by $AdS_5$ backgrounds, observe that if $\sigma_\pm$ are Killing spinors, then
\begin{eqnarray}
\tau_\pm=\Gamma_z \Gamma_x \sigma_\pm~,~~~\tau_\pm=\Gamma_z \Gamma_y \sigma_\pm~,
\end{eqnarray}
are also Killing spinors, and vice versa. As a result if $\sigma_\pm$ are Killing spinors, then  $\sigma'_\pm=\Gamma_{xy} \sigma_\pm$ are also
Killing spinors and similarly for $\tau_\pm$.  As a result $\mathrm{dim}\, \mathrm {Ker}(\nabla^{(\pm)}, {\cal A}^{(\pm)}, {\cal B}^{(\pm)})$
and $\mathrm{dim}\, \mathrm {Ker}(\nabla^{(\pm)}, {\cal A}^{(\pm)}, {\cal C}^{(\pm)})$ are even numbers.

Furthermore, as in the previous cases,   if either $\sigma_-$ or $\tau_-$ is a solution, so is
\begin{eqnarray}
\sigma_+=A^{-1} \Gamma_+ \Gamma_z \sigma_-~,~~~\tau_+=A^{-1} \Gamma_+ \Gamma_z \tau_-~,~~~
\label{pmrelads5}
\end{eqnarray}
and similarly, if either  $\sigma_+$ or  $\tau_+$ is a solution, so is
\begin{eqnarray}
\sigma_-=A \Gamma_- \Gamma_z \sigma_+~,~~~\tau_-=A \Gamma_- \Gamma_z \tau_+~.~~~
\label{mprelads5}
\end{eqnarray}
From the above relations one concludes that the $AdS_5\times_w M^{5}$ backgrounds preserve
\begin{eqnarray}
N= 4 \,\,\mathrm{dim}\, \mathrm {Ker}(\nabla^{(\pm)}, {\cal A}^{(\pm)}, {\cal B}^{(\pm)})= 4 \,\,\mathrm{dim}\, \mathrm {Ker}(\nabla^{(\pm)}, {\cal A}^{(\pm)}, {\cal C}^{(\pm)})=8 k~,
\label{nads5}
\end{eqnarray}
for either $+$  or $-$  choice of sign and $k\in {\mathbb{N}}_{+}$. This confirms (\ref{susyc}) for the $AdS_5$ backgrounds. Of course $N\leq 32$, and for $N=32$
the solutions are locally isometric \cite{n32} to $AdS_5\times S^5$.

\section{$AdS_5$:  Global analysis}

\subsection{A Lichnerowicz type theorem for $\tau_{\pm}$ and $\sigma_{\pm}$}

To extend formula (\ref{indexf}) to $AdS_5$ backgrounds, we shall again prove  a Lichnerowicz type theorem which relates the Killing spinors
to the zero modes of Dirac-like of operators on $M^5$ coupled to fluxes.  The proof is similar to that we have presented in previous cases and so
we shall be brief. It suffices to prove the the Lichnerowicz type of theorem for $\sigma_+$ and $\tau_+$ spinors as the proof for the other pair $\sigma_-$ and $\tau_-$
follows because of the relations (\ref{pmrelads5}) and (\ref{mprelads5}) and the fact that these isomorphisms commute with the relevant operators.

To begin the proof, let us denote  both $\sigma_+$ and $\tau_+$ spinors  collectively with $\chi_+$  and define
\begin{eqnarray}
\mathbb{D}_i^{(+)}= \nabla^{(+)}_i+ q \Gamma_{zi} A^{-1} \mathbb{B}^{(+)}~,
\end{eqnarray}
where
\begin{eqnarray}
\mathbb{B}^{(+)}=-{c\over 2\ell}-{1\over 2} \Gamma_z \slashed\partial A  - {i\over 2} A Y \Gamma_{yx}+{1\over 96} A \Gamma_z \slashed H C*~,
\end{eqnarray}
and $c=1$ when acting on $\sigma_+$ and $c=-1$ when acting on $\tau_+$, ie either $\mathbb{B}^{(+)}={\cal B}^{(+)}$ or $\mathbb{B}^{(+)}={\cal C}^{(+)}$, respectively.
  It is clear
from this that if $\chi_+$ is a Killing spinor, then it is parallel with respect to $\mathbb{D}$.

The modified Dirac-like operator on $M^5$ is
\begin{eqnarray}
\mathscr{D}^{(+)}\equiv \Gamma^i \mathbb{D}_i^{(+)}=\Gamma^i \nabla_i+\Sigma^{(+)}~,
\end{eqnarray}
where
\begin{eqnarray}
\Sigma^{(+)}=  \frac{5 q c}{2\ell } A^{-1}\Gamma_z + \frac{1 + 5 q}{4 } \slashed{\partial} \log A^2 - \frac{i}{2} \slashed{Q} + \frac{5 i - 5 i q}{2} Y \Gamma_{z x y} + \frac{7 - 5 q}{96} \slashed{H} C *~.
\nonumber \\
\end{eqnarray}
Next suppose that $\chi_+$ is a zero mode of  $\mathscr{D}^{(+)}$, ie $\mathscr{D}^{(+)}\chi_+=0$. Then after some Clifford algebra
 computation, which has been presented in appendix \ref{clifads5}, $q=3/5$,  and the use of field equations, one can establish the identity
\begin{eqnarray}
 {\nabla}^2 \left\| \chi_+ \right\| ^2 &+& 5 A^{-1} \partial^{i} A \partial_{i} \left\| \chi_+ \right\| ^2 = 2 \left\| \mathbb{D}^{( + )} \chi_+ \right\| ^2
 \cr
 &&+ \frac{48}{5} A^{-2} \left\| \mathbb{B}^{(+)} \chi_+ \right\| ^2 + \left\| \cal{A}^{(+)} \chi_+ \right\| ^2~.
 \label{ads5maxpp}
\end{eqnarray}
Assuming that  the Hopf maximum principle applies, eg for $M^5$  compact and smooth fields,  the solution of the above equation
reveals that $\chi_+$ is a Killing spinor and that  $\parallel\chi_+\parallel=\mathrm{const}$.

A similar formula to (\ref{ads5maxpp}) can be established for $\sigma_-$ and $\tau_-$ spinors. In particular, we define
\begin{eqnarray}
\mathscr{D}^{(-)}=\Gamma^i \nabla_i+ \Sigma^{(-)}~,
\end{eqnarray}
and
\begin{eqnarray}
\Sigma^{(-)}=- \frac{5 q c}{2\ell } A^{-1}\Gamma_z + \frac{-1+5q }{4} \slashed{\partial} \log A^2 - \frac{i}{2} \slashed{Q} - \frac{5 i - 5 i q}{2} Y \Gamma_{zxy}  + \frac{7 -  5q}{96} \slashed{H} C *~,
\nonumber \\
\end{eqnarray}
where $c=1$ for the $\sigma_-$ spinors while $c=-1$ for $\tau _-$ spinors. As has already been mentioned, because  the relations (\ref{pmrelads5}) and (\ref{mprelads5}) between the
$\sigma_-, \tau_-$ and $\sigma_+, \tau_+$ spinors  commute  with the KSEs and the modified Dirac-like operators, it is not necessary to
prove the maximum principle independently   for $\sigma_-, \tau_-$.
To summarize, we have shown that
\begin{eqnarray}
\nabla^{(\pm)}_i\sigma_\pm=0~,~~{\cal B}^{(\pm)}\sigma_\pm=0~,~~{\cal A}^{(\pm)}\sigma_\pm=0 \Longleftrightarrow \mathscr{D}^{(\pm)}\sigma_\pm=0~;~c=1~,
\cr
\nabla^{(+)}_i\tau_\pm=0~,~~{\cal C}^{(\pm)}\tau_\pm=0~,~~{\cal A}^{(\pm)}\tau_\pm=0 \Longleftrightarrow \mathscr{D}^{(\pm)}\tau_\pm=0~;~c=-1~,
\end{eqnarray}
and that
\begin{eqnarray}
\parallel\sigma_+\parallel&=&\mathrm{const}~,~~~\parallel\tau_+\parallel=\mathrm{const}~,
\cr
A^{-2}\, \parallel\sigma_-\parallel^2&=&  \mathrm{const}~,~~~A^{-2}\,\parallel\tau_-\parallel^2= \mathrm{const}~.
\end{eqnarray}

\subsection{Counting supersymmetries again}

To establish (\ref{indexf}) for $AdS_5$ backgrounds, observe that the dimension of the kernel of $\mathscr{D}^{(\pm)}$ operators is even.  This is because
if $\sigma_\pm$ or $\tau_\pm$ are in the kernel, then  $\Gamma_{xy}\sigma_\pm$ or $\Gamma_{xy}\tau_\pm$ are also
in the kernel.  Since  $\Gamma_{xy}\sigma_\pm$ or $\Gamma_{xy}\tau_\pm$ are linearly independent of $\sigma_\pm$ and $\tau_\pm$,
the dimension of the  kernel of $\mathscr{D}^{(\pm)}$  is an even number.

Next, provided that the data satisfy the requirements
of Hopf maximum principle, we have that
\begin{eqnarray}
N= 4 \,\,\mathrm{dim}\, \mathrm {Ker}(\nabla^{(-)}, {\cal A}^{(-)}, {\cal B}^{(-)})=4\,\, \mathrm{dim}\, \mathrm {Ker} \mathscr{D}^{(-)}_{c=1}~,
\end{eqnarray}
which applies to $\sigma_-$ spinors and  confirms (\ref{indexf}).  A similar formula is valid for the three other choices of spinors.

\section{$AdS_6$: Local analysis}

\subsection{Fields, Bianchi identities and field equations}

For $AdS_p$, $p\geq 6$, the only non-vanishing fluxes are those of the magnetic components of the various field strengths.  Since $F$ is self-dual, $F=0$
for all such backgrounds.   The fields on the horizon section ${\cal S}$ for $AdS_6$ backgrounds are
\begin{eqnarray}
ds^2({\cal S})= A^2( dz^2+e^{2z\over\ell} \sum_{a=1}^3 (dx^a)^2)+ ds^2(M^4)~,~~~G=H~,~~P=\xi~,
\end{eqnarray}
and $h=-{2\over \ell} dz-2 d\log A$ and $\Delta=X=L=0$, where $x^1=x, x^2=y$ as for $AdS_5$ and $x^3=w$.

Substituting the above fields into the Bianchi identities (\ref{ba0}) and (\ref{ba}), we find
\begin{eqnarray}
 dH = i Q \wedge H - \xi \wedge \overline{H}~,~~~
 d\xi =2 i Q \wedge \xi~,~~~ dQ = -i \xi \wedge \overline{\xi} .
\end{eqnarray}
 Similarly, the field equations (\ref{feq1})-(\ref{feq7}) give
\begin{eqnarray}
 {\nabla}^{i} H_{i j k} &=& -6 \partial^i \log A \, H_{i j k} + i Q^{i} H_{i j k} + \xi^{i} \overline{H}_{i j k}~,
 \cr
 {\nabla}^{i} \xi_{i} &=& -6 \partial^i \log A \, \xi_i + 2 i Q^{i} \xi_{i} - \frac{1}{24} H^2~,
\cr
 A^{-1} {\nabla}^2 A &=&  \frac{1}{48} \parallel H \parallel^2 - \frac{5}{\ell^2} A^{-2} - 5 (d\log A)^2~,
\cr
  {R}^{(4)}_{i j}&=&  6 A^{-1} {\nabla}_{i} {\nabla}_{j} A
  \cr
 &+& \frac{1}{4} H_{( i}{}^{k \ell} \overline{H}_{j ) k \ell} - \frac{1}{48} \parallel H \parallel^2 \delta_{i j} + 2 \xi_{( i} \overline{\xi}_{j )}~.
 \label{ads6feq}
\end{eqnarray}
This concludes the analysis of Bianchi identities and field equations.

\subsubsection{The warp factor is nowhere vanishing}

As in the previous AdS backgrounds, one can show that the warp factor $A$ is no-where vanishing. The argument is based on the third field equation in (\ref{ads6feq}).

\subsection{Solution of KSEs}

The solution of the spatial horizon section ${\cal S}$ KSEs   (\ref{horkses}) reveals that
\begin{eqnarray}
\eta_+=\sigma_+-{1\over\ell} (\sum_a x^a \Gamma_a) \Gamma_z \tau_++ e^{-{z\over \ell}} \tau_+~,~~\eta_-=\sigma_-+ e^{{z\over\ell}} \big(-{1\over\ell}
(\sum_a x^a \Gamma_a) \Gamma_z \sigma_-+\tau_-\big)~,
\nonumber \\
\end{eqnarray}
where $\sigma_\pm$ and $\tau_\pm$ depend only on the coordinates of $M^4$. After taking into account all the integrability conditions, the remaining independent KSEs are
\begin{eqnarray}
&&\nabla_i^{(\pm)}\sigma_\pm=0~,~~~\nabla_i^{(\pm)}\tau_\pm=0~,~~~{\cal A}^{(\pm)}\sigma_\pm=0~,~~~{\cal A}^{(\pm)}\tau_\pm=0~,
\cr
&&{\cal B}^\pm\sigma_\pm=0~,~~~{\cal C}^\pm\tau_\pm=0~,
\end{eqnarray}
where
\begin{eqnarray}
&&\nabla_i^{(\pm)}=  \nabla_i+ \Psi^{(\pm)}_i~,~~~{\cal A}^{(\pm)}={1\over24} \slashed H+\slashed\xi C*~,
\cr
&&{\cal B}^{(\pm)}=\Xi_\pm~,~~~{\cal C}^{(\pm)}=\Xi_\pm\pm {1\over\ell}~,
\end{eqnarray}
and
\begin{eqnarray}
\Psi_i^{(\pm)}&=&\pm{1\over2} \partial_i\log A-{i\over2} Q_i+ \big(-{1\over96} (\Gamma\slashed H)_i+{3\over32} \slashed H_i\big) C*~,
\cr
\Xi_\pm&=&\mp {1\over 2\ell}-{1\over2} \Gamma_z \partial_i A \Gamma^i+ {1\over 96} A \Gamma_z \slashed H C*~.
\end{eqnarray}
This concludes the solution of the KSEs along the $AdS_6$ directions.

\subsection{Counting supersymmetries}

A direct inspection of the KSEs reveals that  if $\sigma_\pm$ are Killing spinors, then
\begin{eqnarray}
\tau_\pm=\Gamma_z \Gamma_a \sigma_\pm~,
\end{eqnarray}
are also Killing spinors, and vice versa. As a result if $\sigma_\pm$ are Killing spinors, then  $\sigma'_\pm=\Gamma_{ab} \sigma_\pm$ are also
Killing spinors and similarly for $\tau_\pm$.  Therefore $\mathrm{dim}\, \mathrm {Ker}(\nabla^{(\pm)}, {\cal A}^{(\pm)}, {\cal B}^{(\pm)})$
and $\mathrm{dim}\, \mathrm {Ker}(\nabla^{(\pm)}, {\cal A}^{(\pm)}, {\cal C}^{(\pm)})$ are multiples of four.

Furthermore, as in the previous cases,   if either $\sigma_-$ or $\tau_-$ is a solution, so is
\begin{eqnarray}
\sigma_+=A^{-1} \Gamma_+ \Gamma_z \sigma_-~,~~~\tau_+=A^{-1} \Gamma_+ \Gamma_z \tau_-~,~~~
\label{pmrelads6}
\end{eqnarray}
and similarly, if either  $\sigma_+$ or  $\tau_+$ is a solution, so is
\begin{eqnarray}
\sigma_-=A \Gamma_- \Gamma_z \sigma_+~,~~~\tau_-=A \Gamma_- \Gamma_z \tau_+~.~~~
\label{mprelads6}
\end{eqnarray}
From the above relations one concludes that the $AdS_5\times_w M^{5}$ backgrounds preserve
\begin{eqnarray}
N= 4 \,\,\mathrm{dim}\, \mathrm {Ker}(\nabla^{(\pm)}, {\cal A}^{(\pm)}, {\cal B}^{(\pm)})= 4 \,\,\mathrm{dim}\, \mathrm {Ker}(\nabla^{(\pm)}, {\cal A}^{(\pm)}, {\cal C}^{(\pm)})=16 k~,
\nonumber \\
\label{nads5}
\end{eqnarray}
for either $+$  or $-$  choice of sign and $k\in {\mathbb{N}}_{+}$. This confirms (\ref{susyc}) for the $AdS_6$ backgrounds. It turns out that there can be
$AdS_6$ backgrounds for only  $N=16$ as  there are no such backgrounds preserving $N=32$ supersymmetries \cite{n32}.

\section{$AdS_6$:  Global analysis}

\subsection{A Lichnerowicz type theorem for $\tau_{\pm}$ and $\sigma_{\pm}$}

As in previous cases, let us prove a Lichnerowicz type theorem for $\sigma_+$ and $\tau_+$ spinors. For this denote $\sigma_+$ and $\tau_+$ collectively by $\chi_+$ and define
\begin{eqnarray}
\mathbb{D}_i^{(+)}= \nabla^{(+)}_i+ q \Gamma_{zi} A^{-1} \mathbb{B}^{(+)}~,
\end{eqnarray}
where
\begin{eqnarray}
\mathbb{B}^{(+)}=-{c\over 2\ell}-{1\over 2} \Gamma_z \slashed\partial A \Gamma^i +{1\over 96} A \Gamma_z \slashed H C*~,
\end{eqnarray}
and $c=1$ when acting on $\sigma_+$ and $c=-1$ when acting on $\tau_+$, ie either $\mathbb{B}^{(+)}={\cal B}^{(+)}$ or $\mathbb{B}^{(+)}={\cal C}^{(+)}$, respectively.
  It is clear
from this that if $\chi_+$ is a Killing spinor, then it is parallel with respect to $\mathbb{D}$.

The modified Dirac-like operator on $M^4$ is
\begin{eqnarray}
\mathscr{D}^{(+)}\equiv \Gamma^i \mathbb{D}_i^{(+)}=\Gamma^i \nabla_i+\Sigma^{(+)}~,
\end{eqnarray}
where
\begin{eqnarray}
\Sigma^{(+)}= \frac {2 q c}{\ell} A^{-1} \Gamma_z + \frac{1 + 4 q}{4 } \slashed{\partial} \log A^2 - \frac{i}{2} \slashed{Q} + \frac{8 - 4 q}{96} \slashed{H} C *~.
\end{eqnarray}
Next suppose that $\chi_+$ is a zero mode of  $\mathscr{D}^{(+)}$, ie $\mathscr{D}^{(+)}\chi_+=0$. Then after some Clifford algebra
 computation, which has been presented in appendix \ref{clifads6}, $q=1$,  and the use of field equations, one can establish the identity
\begin{eqnarray}
 {\nabla}^2 \left\| \chi_+ \right\| ^2 &+& 6 A^{-1} \partial^{i} A \partial_{i} \left\| \chi_+ \right\| ^2 = 2 \left\| \mathbb{D}^{( + )} \chi_+ \right\| ^2
 \cr
 &&+ 16 A^{-2} \left\| \mathbb{B}^{(+)} \chi_+ \right\| ^2 + \left\| \cal{A}^{(+)} \chi_+ \right\| ^2~.
 \label{ads6maxpp}
\end{eqnarray}
Assuming that  the Hopf maximum principle applies, eg for $M^4$  compact and smooth fields,  the solution of the above equation
reveals that $\chi_+$ is a Killing spinor and that  $\parallel\chi_+\parallel=\mathrm{const}$.

A similar formula to (\ref{ads5maxpp}) can be established for $\sigma_-$ and $\tau_-$ spinors. In particular, we define
\begin{eqnarray}
\mathscr{D}^{(-)}=\Gamma^i \nabla_i+ \Sigma^{(-)}~,
\end{eqnarray}
and
\begin{eqnarray}
\Sigma^{(-)}=- \frac{2 q c}{ \ell} A^{-1}\Gamma_z + \frac{-1+4q }{4} \slashed{\partial} \log A^2 - \frac{i}{2} \slashed{Q}   + \frac{8 -  4q}{96} \slashed{H} C *~,
\end{eqnarray}
where $c=1$ for the $\sigma_-$ spinors while $c=-1$ for $\tau _-$ spinors. Because of the relations (\ref{pmrelads5}) and (\ref{mprelads5}) between the
$\sigma_-, \tau_-$ and $\sigma_+, \tau_+$ spinors and the commutation of these relations with the KSEs and the associated Dirac-like operators, it is not necessary to
prove the maximum principle independently   for $\sigma_-, \tau_-$.
To summarize, we have shown that
\begin{eqnarray}
\nabla^{(\pm)}_i\sigma_\pm=0~,~~{\cal B}^{(\pm)}\sigma_\pm=0~,~~{\cal A}^{(\pm)}\sigma_\pm=0 \Longleftrightarrow \mathscr{D}^{(\pm)}\sigma_\pm=0~;~c=1~,
\cr
\nabla^{(+)}_i\tau_\pm=0~,~~{\cal C}^{(\pm)}\tau_\pm=0~,~~{\cal A}^{(\pm)}\tau_\pm=0 \Longleftrightarrow \mathscr{D}^{(\pm)}\tau_\pm=0~;~c=-1~,
\end{eqnarray}
and that
\begin{eqnarray}
\parallel\sigma_+\parallel&=&\mathrm{const}~,~~~\parallel\tau_+\parallel=\mathrm{const}~,
\cr
A^{-2}\, \parallel\sigma_-\parallel^2&=&  \mathrm{const}~,~~~A^{-2}\,\parallel\tau_-\parallel^2= \mathrm{const}~.
\end{eqnarray}

\subsection{Counting supersymmetries again}

To establish (\ref{indexf}) for $AdS_5$ backgrounds, observe that the dimension of the kernel of $\mathscr{D}^{(\pm)}$ operators is multiple of 4.  This is because
if $\sigma_\pm$ or $\tau_\pm$ are in the kernel, then  $\Gamma_{ab}\sigma_\pm$ or $\Gamma_{ab}\tau_\pm$ are also
in the kernel.  Since  $\Gamma_{ab}\sigma_\pm$ or $\Gamma_{ab}\tau_\pm$ are linearly independent of $\sigma_\pm$ and
$\tau_\pm$,
the dimension of the  kernel of $\mathscr{D}^{(\pm)}$  is $4k$.

Next provided that the data satisfy the requirements
of Hopf maximum principle, we have that
\begin{eqnarray}
N= 4 \,\,\mathrm{dim}\, \mathrm {Ker}(\nabla^{(-)}, {\cal A}^{(-)}, {\cal B}^{(-)})=4\,\, \mathrm{dim}\, \mathrm {Ker} \mathscr{D}^{(-)}_{c=1}=16k~,
\end{eqnarray}
which applies to $\sigma_-$ spinors and  confirms (\ref{indexf}).  A similar formula is valid for the three other choices of spinors.

\section{AdS$_n$, for $ n \geq 7 $}

There are no supersymmetric AdS$_n$, $ n \geq 7 $ IIB backgrounds, see also \cite{iibadsf} where this result has been established assuming that the Killing spinors factorize.
To see this first observe that  if a background preserves at least one supersymmetry, then the three-form, $ H $, is zero. For AdS$_n$, $ n \geq 8 $, this is automatically true. For AdS$_7$, we can show this  by manipulating the algebraic Killing spinor equation,
\begin{equation}
 \left( \slashed{\xi} C * + \frac{1}{24} \slashed{H} \right) \sigma_+ = 0 .
\end{equation}

We start by multiplying this by $ \slashed{\overline{H}} $ to convert it to an eigenvalue equation,
\begin{equation}
 \slashed{\xi} \slashed{\overline{H}} C * \sigma_+ = -\frac{1}{4} \parallel H \parallel^2 \sigma_+ ,
\end{equation}
and then we square the operator on the left hand side to eliminate $ C * $,
\begin{equation}
 \xi_i \overline{\xi}_j \Gamma^{i j} \sigma_+ = -\left( \parallel \xi \parallel^2 + \frac{1}{96} \parallel H \parallel^2 \right) \sigma_+ .
\end{equation}
Finally, squaring this operator as well, we end up with a scalar equation
\begin{equation}
 \parallel\xi \parallel^4 -\parallel\xi^2\parallel^2 = \left( \parallel\xi \parallel^2 + \frac{1}{96} \parallel H \parallel^2 \right) ^2 ,
\end{equation}
from which we conclude that $ \xi^2 = \xi_i \xi^i $ and $H$ are both zero.

Having shown that $ H = 0 $, the integrability condition, \eqref{alg1}, reduces to
\begin{equation}
 \left( \frac{1}{4 \ell^2} + \frac{1}{4}( dA )^2 \right) \sigma_+ = 0 ,
\end{equation}
which has no solution. Therefore, there are no supersymmetric AdS$_n$ backgrounds for $ n \geq 7 $.

\begin{table}
\centering
\fontencoding{OML}\fontfamily{cmm}\fontseries{m}\fontshape{it}\selectfont
\begin{tabular}{|c|c|}\hline
$AdS_n\times_w M^{10-n}$& $N$
 \\
\hline\hline

$n=2$&$2k, k<14$
\\
\hline
$n=3$&$2k, k<14$
\\
\hline
$n=4$&$4k, k<7$
\\
\hline
$n=5$&$8k, k\leq 4$
\\
\hline
$n=6$&$16$
\\
\hline
$n\geq 7$&$-$
\\
\hline
\end{tabular}
\label{tab2}
\begin{caption}
{\small {\rm ~~The number of supersymmetries $N$ of $AdS_n\times_w M^{10-n}$ backgrounds are given. For $AdS_2\times_w M^{8}$, one can show that these backgrounds preserve an even number of supersymmetries
provided that they are smooth and $M^8$ is compact without boundary. For the rest, the counting of supersymmetries does not rely on the compactness of $M^{10-n}$. The bounds in $k$ arise from the non-existence
 of supersymmetric solutions with near maximal and maximal supersymmetry. For the remaining fractions, it is not known whether there always exist backgrounds preserving the prescribed
 number of supersymmetries. Supersymmetric $AdS_n$, $n\geq 7$, backgrounds do not exist.
}}
\end{caption}
\end{table}

\section{Flat IIB backgrounds}

Warped flat backgrounds ${\mathbb{R}}^{n-1, 1}\times_w M^{10-n}$ are also included in our analysis.  These arise in the ``flat limit'', ie the limit  that the $AdS_n$ radius $\ell$ is taken to infinity.
This limit is smooth in all our computations.  However, some of our results on $AdS_n$ backgrounds do not extend to the flat backgrounds.  The investigation
of the KSEs is also somewhat different from that of $AdS_n$ backgrounds.

To emphasize some of the differences between $AdS_n$ and ${\mathbb{R}}^{n-1, 1}$ backgrounds, it has been known for sometime that there are no smooth warped flux compactifications in  supergravity \cite{maldacena2}. To alter this either additional sources have to be added to the
supergravity equations, like brane charges,  and /or consider higher order curvature  corrections which arise for example from anomaly cancellation mechanisms or $\alpha'$ corrections in string theory. In either case,
the new backgrounds can be constructed  as  corrections  to supergravity solutions. Because there are different sources that can be added and we do not have control
over all higher curvature corrections, we shall mostly focus here on the supergravity limit and explore the similarities and differences between the $AdS_n$ and ${\mathbb{R}}^{n-1, 1}$ backgrounds.

\subsection{The warp factor is not nowhere vanishing}

We have seen that the warp factor in all $AdS_n$ is no-where vanishing.  This does not extend to ${\mathbb{R}}^{n-1, 1}$ backgrounds because the finiteness of the
$AdS_n$ radius has been essential in the proof of the statement. In fact $A$ must vanish somewhere for non-trivial  ${\mathbb{R}}^{n-1, 1}$ backgrounds with fluxes. This follows from
the results of  \cite{maldacena2} on the non-existence of smooth warped flux compactifications in the context of supergravity. To see this,
let us focus on the ${\mathbb{R}}^{1,1}$ case, as the argument is similar in all the other cases. If $A$ is no-where vanishing and $M^8$ is compact,
an application of the maximum principle on the field equation for $A$ (\ref{ads2feq5}) reveals that $A$ is constant and  the fluxes $F$ and $G$
vanish.  Furthermore using the formula
\begin{eqnarray}
\nabla^2 \parallel \xi\parallel^2=2 (\nabla_{(i} \xi_{j)}-2i \Lambda_{(i} \xi_{j)}) (\nabla^{(i} \xi^{j)}-2i \Lambda^{(i} \xi^{j)})+6 (\parallel \xi\parallel^2)^2~,
\end{eqnarray}
established in \cite{iibhor} and upon using again the maximum principle, one can show that $\xi=0$.  As a result all the form field strengths vanish
which is a contradiction.  From now on, we shall assume that $A$ is non-vanishing on some dense subset of $M^{10-n}$ and carry out the analysis
that follows on that subset.

\subsection{Counting supersymmetries} \label{fcountf}

All the local computations we have done for $AdS_n$ backgrounds extend to ${\mathbb{R}}^{n-1, 1}$ backgrounds. However the statements which rely
on the smoothness of the fields as well as the non-vanishing of the warp factor have to be re-examined. In particular, the solution
of the KSEs can be carried out as has been described for $AdS_n$. Also the various maximum principle formulae are valid away from points where $A=0$, like eg (\ref{lichident}),
(\ref{lichident2}), (\ref{ads3maxp}) and others.  However, the Hopf maximum principle cannot be applied any longer even if $M^{10-n}$ is taken to be compact.
As a result there is not a straightforward relation between Killing spinors and zero modes on Dirac-like operators on $M^{10-n}$.
Because of this,  for the counting of supersymmetries we shall rely on the local solution of the KSEs as presented for the $AdS_n$ backgrounds.

\subsubsection{${\mathbb{R}}^{1,1}$ backgrounds}

The counting of supersymmetries for $AdS_2$ backgrounds relies on the global properties of $M^8$ and the smoothness of the fields.  As a result,
the number of supersymmetries preserved by  ${\mathbb{R}}^{1,1}$ backgrounds cannot be concluded. In particular, it is not apparent that such backgrounds always preserve an
even number of supersymmetries.  Nevertheless, if $\eta_-$ is a Killing spinor, so is $\Gamma_+\Theta_-\eta_-$ on $M^8$. Now if $\mathrm{Ker}~\Theta_-=\{0\}$,
it is clear that there will be a doubling of supersymmetries.  In such a case, the number of Killing spinors for such backgrounds is $N\geq 2N_-$, where $N_-$ is the number of $\eta_-$ Killing spinors.

\subsubsection{${\mathbb{R}}^{2,1}$ backgrounds}

Let us re-examine the solution of the KSEs. In the limit $\ell\rightarrow \infty$, the integrability conditions
(\ref{alg1}) become
\begin{eqnarray}
\Theta_\mp \Theta_\pm \eta_\pm=0~.
\label{algads31}
\end{eqnarray}
In the same limit, the solution of the KSEs (\ref{zads3})  along the z-direction is
\begin{eqnarray}
\eta_\pm=\sigma_\pm+ z \Xi_\pm \tau_\pm~,~~~\Xi_\pm(\sigma_\pm-\tau_\pm)=0~,
\end{eqnarray}
where $\Xi_\pm=A \Gamma_z \Theta_\pm$. The integrability conditions are  automatically satisfied because of (\ref{algads31}). The remaining independent KSEs
are
\begin{eqnarray}
&&\nabla^{(\pm)}_i \sigma_\pm=0~,~~~\nabla^{(\pm)}_i\tau_\pm=0~,
\cr
&&{\cal A}^{(\pm)} \sigma_\pm=0~,~~~{\cal A}^{(\pm)} \tau_\pm=0~,
\end{eqnarray}
where $\nabla^{(\pm)}$ and ${\cal A}^{(\pm)}$ are given in (\ref{ppalgads3}). $\tau_\pm$ and $\sigma_\pm$ satisfy the same  differential equations and are not
linearly independent. As a result, it suffices to consider only the $\sigma_\pm$ spinors and set $\tau_\pm=\sigma_\pm$.  Therefore the number of supersymmetries
preserved by ${\mathbb{R}}^{2,1}$ backgrounds is $N=\mathrm{dim} \mathrm{Ker} (\nabla^{(+)}, {\cal A}^{(+)})+\mathrm{dim} \mathrm{Ker} (\nabla^{(-)}, {\cal A}^{(-)})$.

Next, it is straightforward to observe that if $\sigma_-$ is a solution of (\ref{ads3indkse}) in the limit $\ell=\infty$, then
\begin{eqnarray}
\sigma_+=A^{-1} \Gamma_z\Gamma_+\sigma_-~,~~~
\end{eqnarray}
is also a solution.  Conversely, if  $\sigma_+$ is a solution,
then
\begin{eqnarray}
\sigma_-=A \Gamma_z\Gamma_-\sigma_+~,~~~
\end{eqnarray}
is also a solution.  Therefore, $\mathrm{dim} \mathrm{Ker} (\nabla^{(+)}, {\cal A}^{(+)})=\mathrm{dim} \mathrm{Ker} (\nabla^{(-)}, {\cal A}^{(-)})$, and so the ${\mathbb{R}}^{2,1}$ backgrounds preserve an even number of supersymmetries.

Observe that in general the Killing spinors can depend non-trivially on the $z$ coordinate.  This is possible only if $\sigma_\pm\notin \mathrm{Ker}\, \Xi_\pm$
even though it is required that  $\sigma_\pm\in \mathrm{Ker}\, \Xi^2_\pm$ because of (\ref{algads31}).

\subsubsection{${\mathbb{R}}^{3,1}$ backgrounds}

The counting of supersymmetries of ${\mathbb{R}}^{3,1}$ backgrounds is similar to ${\mathbb{R}}^{2,1}$ solutions. In particular integrating the KSEs along the $z$ and $x$ directions
we find that
\begin{eqnarray}
\eta_\pm= \sigma_\pm+ A (z \Gamma_z+ x\Gamma_x) \Theta_\pm \tau_\pm~,~~~\Theta_\pm(\sigma_\pm-\tau_\pm)=0~,
\end{eqnarray}
with $\sigma_\pm$ and $\tau_\pm$ both in the kernel of $(\nabla^{(\pm)}, {\cal A}^{(\pm)})$ given in (\ref{ppalgads4}) and $\Xi_\pm=A \Gamma_z \Theta_\pm$.  Therefore as in the ${\mathbb{R}}^{2,1}$ case these spinors are not linearly independent
and so suffices to consider $\sigma_\pm$ and set $\tau_\pm=\sigma_\pm$.
 In addition if $\sigma_+$ is a solution, so is $\Gamma_{zx} \sigma_+$.  This together with the fact that if $\sigma_+$ is a solution so is $\sigma_-=A \Gamma_z\Gamma_-\sigma_+$, and vice versa if $\sigma_-$
 is a solution so is $\sigma_+=A^{-1} \Gamma_z\Gamma_+\sigma_-$, one concludes that  ${\mathbb{R}}^{3,1}$ backgrounds preserve $N=4k$ supersymmetries.

 Note again that the Killing spinors are allowed to depend linearly on the coordinates of ${\mathbb{R}}^{3,1}$.  This is the case only if $\tau_\pm\notin \mathrm{Ker}\, \Xi_\pm$
 even though it is required that $\tau_\pm\in \mathrm{Ker}\, \Xi^2_\pm$ because of (\ref{algads31}).

\subsubsection{${\mathbb{R}}^{n-1,1}$, $n>4$,  backgrounds}

As in the previous cases, one can prove that
\begin{eqnarray}
\eta_\pm=\sigma_\pm+ A (\sum_{\mu} x^\mu \Gamma_\mu)\Theta_\pm \tau_\pm~,~~~\Theta_\pm(\sigma_\pm-\tau_\pm)=0~,
\end{eqnarray}
 in the limit $\ell \rightarrow \infty$, and that the only linearly independent Killing spinors are $\sigma_\pm$, where $x^\mu$ are all the coordinates of ${\mathbb{R}}^{n-1,1}$ apart from the lightcone ones $u,r$.  Moreover,
 it suffices to count the linearly independent $\sigma_+$ spinors as the $\sigma_-$ spinors can be constructed as $\sigma_-=A \Gamma_z\Gamma_-\sigma_+$ from the $\sigma_+$ ones, and vice versa
 because of the relation $\sigma_+=A^{-1} \Gamma_z\Gamma_+\sigma_-$.

 Next given a $\sigma_+$ Killing spinor, one can see by direct inspection of the KSEs on $M^{10-n}$ that $\Gamma_{ab} \sigma_+$, $a<b$, are also Killing spinors, where $\Gamma_a$ are the gamma matrices
 in directions orthogonal  to $+,-$.  It turns out that for $n=5$, these are all linearly independent and therefore these backgrounds preserve
 $N=8k$  supersymmetries.

 For $n=6$, apart from $\Gamma_{ab} \sigma_+$, $a<b$, observe that  $\Gamma_{a_1a_2a_3a_4}\sigma_+$, $a_1<a_2<a_3<a_4$ also solve the KSEs on $M^4$. However,
 there is a unique Clifford algebra element$\Gamma_{a_1a_2a_3a_4}$, $a_1<a_2<a_3<a_4$, in this case and has eigenvalues $\pm1$, and commutes with all the KSEs. Now if $\sigma_+$ is in one of the two eigenspaces, only four of the
 7 Killing spinors $\{\sigma_+, \Gamma_{ab}\sigma_+\vert a<b\}$ are linearly independent.  Therefore the ${\mathbb{R}}^{5,1}$  backgrounds preserve $N=8k$ supersymmetries.

 Suppose now that $n=7$. Given a Killing spinor $\sigma_+$, then $\Gamma_{ab} \sigma_+$ and $\Gamma_{a_1a_2a_3a_4}\sigma_+$, $a_1<a_2<a_3<a_4$, are also Killing spinors.
There are five $\Gamma_{a_1a_2a_3a_4}$, $a_1<a_2<a_3<a_4$ Clifford algebra operations in this case. Choose one, say $\Gamma_{[4]}$. As in the previous case $\sigma_+$ can be in one of the eigenspaces of $\Gamma_{[4]}$.  In such a case, only
 8 of the previous 16 Killing spinors are linearly independent.  Therefore, the ${\mathbb{R}}^{6,1}$ backgrounds preserve $N=16k$ supersymmetries.
 Of course as a consequence of \cite{n32} the non-trivial  ${\mathbb{R}}^{6,1}$ backgrounds preserve strictly $16$ supersymmetries.  Furthermore adapting the analysis of section 13
 in the limit of infinite AdS radius, one finds that $A$ must be constant, $H=0$ and $\xi_i\xi^i=0$.

 Next take $n=8$.  Given a Killing spinor $\sigma_+$, then $\Gamma_{ab} \sigma_+$,  $\Gamma_{a_1a_2a_3a_4}\sigma_+$, $a_1<a_2<a_3<a_4$, and  $\Gamma_{a_1\dots a_6}\sigma_+$,
 $a_1<\dots<a_6$ are also Killing spinors. All fifteen $\Gamma_{a_1a_2a_3a_4}$, $a_1<a_2<a_3<a_4$, Clifford algebra operators commute with the KSEs and have eigenvalues
 $\pm1$.  Taking a commuting pair of such operators, say $\Gamma_{[4]}$ and $\Gamma'_{[4]}$, and choosing $\sigma_+$ to lie in a common eigenspace of both
 these operators, only eight of the 32 spinors mentioned above are linearly independent.  As a result, ${\mathbb{R}}^{7,1}$ backgrounds preserve $N=16k$ supersymmetries. In fact
 non-trivial  ${\mathbb{R}}^{7,1}$ backgrounds backgrounds, like the D7-brane, preserve strictly 16 supersymmetries. Again for this backgrounds $A$ is constant and $\xi_i\xi^i=0$.
 Furthermore, it can be easily seen from the results of section 13 and after taking the AdS radius to infinity that there are no non-trivial ${\mathbb{R}}^{8,1}$ supersymmetric
  backgrounds.

It should be further noted that  all ${\mathbb{R}}^{n-1,1}$ with $N>16$ are homogeneous spaces \cite{jose}. If one can  show that $A$ is invariant and so constant, then the field equation
of the warp factor implies that there are no such no trivial backgrounds preserving $N>16$ supersymmetries. It is likely that this is the case for all such backgrounds
for which the Killing spinors do not exhibit a  ${\mathbb{R}}^{n-1,1}$ coordinate dependence.

 \begin{table}
\centering
\fontencoding{OML}\fontfamily{cmm}\fontseries{m}\fontshape{it}\selectfont
\begin{tabular}{|c|c|}\hline
${\mathbb{R}}^{n-1,1}\times_w M^{10-n}$& $N$
 \\
\hline\hline

$n=2$&$ N< 28$
\\
\hline
$n=3$&$2k, k<14$
\\
\hline
$n=4$&$4k, k<14$
\\
\hline
$n=5$&$8, 16, 24$
\\
\hline
$n=6$&$8, 16, 24$
\\
\hline
$n= 7$&$16$
\\
\hline
$n= 8$&$16$
\\
\hline
$n= 10$&$32$
\\
\hline
\end{tabular}
\label{tab2}
\begin{caption}
{\small {\rm ~~The number of supersymmetries $N$ of ${\mathbb{R}}^{1,1}\times_w M^{10-n}$ is not a priori an even number. The corresponding statement for $AdS_2$ backgrounds
is proven using global considerations which are not applicable in this case.  For the rest, the counting of supersymmetries follows from the properties of KSEs
 and the classification results of \cite{n31, n29, n32, n28}. All backgrounds with
$n>8$ are maximally supersymmetric and so  locally isometric to ${\mathbb{R}}^{9,1}$.
}}
\end{caption}
\end{table}

\section{On the factorization of Killing spinors}

In many of the investigations of $AdS_n\times M^{10-n}$ backgrounds in IIB and other theories, it is assumed that
the Killing spinors of the spacetime factorize into a product
\begin{eqnarray}
\epsilon=\psi\otimes \chi~,
\label{factorx}
\end{eqnarray}
where $\psi$ is a Killing spinor on the AdS spaces satisfying the equation
\begin{eqnarray}
\nabla_\mu \psi+ \lambda \gamma_\mu \psi=0~,
\label{adsksex}
\end{eqnarray}
and where $\nabla$ and $\gamma_\mu$ are the spin connection and gamma matrices on $AdS_n$, respectively.
Since we have solved the KSEs on the whole spacetime, we can now test this hypothesis. To do this
observe that if the hypothesis is correct, then $\epsilon$ also solves (\ref{adsksex}).
So it suffices to substitute our Killing spinors into (\ref{adsksex}) to see whether
they automatically satisfy it.  This computation is similar that that we have done
for M-theory in \cite{Mads}. It turns out that the Killing spinors $\epsilon$  solve
(\ref{adsksex}) iff
\begin{eqnarray}
\Gamma_z \epsilon=\pm \epsilon~.
\label{extracon}
\end{eqnarray}
However our Killing spinors do {\it not} satisfy this equation. As a result the original hypothesis
is not valid in general.

To illustrate that (\ref{extracon}) is restrictive, we shall test it against the supersymmetry counting
for the $AdS_5\times S^5$ background. It is known that this background preserves all 32 supersymmetries.  It can be easily seen that to solve the
algebraic KSEs for this background in  (\ref{ads5x}) for the $\tau_+$ spinor, one has to impose
\begin{eqnarray}
\Gamma_{xy}\tau_+=\pm i \tau_+~.
\end{eqnarray}
After choosing one of the signs, it is clear that the dimension of the space of solutions is 8 counted over the reals. The gravitino KSE is then solved without any additional
constraints on $\tau_+$.  Next using the relation between $\tau_+$, $\tau_-$, $\sigma_+$ and $\sigma_-$ solutions to the KSEs, we conclude that the number of
Killing spinors of this background is $4\times 8=32$ as expected.  However if one also imposes the condition (\ref{extracon}) on $\tau_+$, one will arrive at the incorrect conclusion
that $AdS_5\times S^5$ preserves only 16 supersymmetries.

We have seen that the spinor factorization assumption   in (\ref{factorx}) leads to the incorrect counting of supersymmetries for AdS backgrounds.
It is also likely that it puts additional restrictions on the geometry of the transverse spaces $M^{10-n}$. We shall investigate this in another
publication.

To continue, let us examine the factorization of the Killing spinors as in (\ref{factorx}) for flat backgrounds to see whether
a similar issue arises as for the AdS. A direct inspection of the Killing spinors we have found in section \ref{fcountf} reveals that
the Killing spinors do not solve the KSEs on ${\mathbb{R}}^{n-1,1}$ whenever they have an explicit dependence on the coordinates of ${\mathbb{R}}^{n-1,1}$.
As we have already stressed, this dependence appears whenever $\sigma_\pm$ are not in the kernel of $\Theta_\pm$. However it is required
as a consequence of the KSEs, field equations and Bianchi identities that $\Theta_\mp \Theta_\pm \sigma_\pm=0$.
Thus assuming that the Killing spinor factorize as in (\ref{factorx}) with $\psi$ to be a constant spinor on ${\mathbb{R}}^{n-1,1}$, we find that this imposes the
additional condition  $\Theta_\pm\sigma_\pm=0$ on the Killing spinors. It is not apparent that this condition always holds for flat backgrounds.
On the other hand we are not aware of examples for which it does not, and so the question will be investigated further elsewhere.

\section{Conclusions}

We have determined the a priori fractions of supersymmetry preserved by the warped $AdS_n$ and flat backgrounds ${\mathbb{R}}^{n-1,1}$ in IIB supergravity.
The results are tabulated in tables 1 and 2, and in equations (\ref{susyc}) and (\ref{susycf}),  respectively.  To achieve this, we have solved the KSEs of IIB supergravity
without making any assumptions on the form of the fields and Killing spinors, and identified the independent KSEs on the transverse spaces $M^{10-n}$. There are two ways to count the number of supersymmetries
for $AdS_n$ backgrounds.  One is directly from the KSEs on $M^{10-n}$ and the other is from counting the zero modes of a suitable
Dirac-like operator on $M^{10-n}$ coupled to fluxes.  For the latter, we have proven new Lichnerowicz type theorems using the Hopf
maximum principle which relates the Killing spinors to the zero modes of the Dirac-like operator.  As a consequence, we have extended
the Lichnerowicz theorem for connections with holonomy contained in a $GL$ group.

The solution of the KSEs of ${\mathbb{R}}^{n-1,1}$ backgrounds can be recovered from that of $AdS_n$ in the limit that the AdS radius goes to infinity.
The counting of supersymmetries for such backgrounds then proceeds by  counting  the solutions of the KSEs on $M^{10-n}$.  The counting
of Killing spinors for ${\mathbb{R}}^{n-1,1}$ backgrounds is different from that of $AdS_n$ backgrounds because of differences in
the identification of linearly independent Killing spinors. Furthermore,  unlike
the $AdS_n$ case,  ${\mathbb{R}}^{n-1,1}$ backgrounds do not satisfy the regularity assumptions of $AdS_n$ backgrounds and so there is no
corresponding counting of supersymmetries via the counting of zero modes of  Dirac-like operators.

Our result is  the first step towards the classification of all  $AdS_n$ and flat backgrounds ${\mathbb{R}}^{n-1,1}$ in IIB supergravity.
The next step is to investigate the existence of backgrounds for each fraction of supersymmetry preserved. We have already excluded
the existence of many cases as can be seen in tables 1 and 2.  However, it is likely that further cases can be excluded especially in the $AdS_n$ case after additional
conditions are put on the transverse space $M^{10-n}$ like for example compactness. The exploration of this question as well as
the geometry of all $AdS_n$ backgrounds will be presented elsewhere.

\vskip 0.3cm

\noindent{\bf Acknowledgements} \vskip 0.1cm
GP is partially supported by the STFC grant ST/J002798/1.
JG is supported by the STFC grant, ST/1004874/1.
JG would like to thank the
Department of Mathematical Sciences, University of Liverpool for hospitality during which part of this work
was completed.

\vskip 0.5cm

\newpage

\appendix

\section{Conventions}

Our form conventions are as follows. Let $\omega$ be a k-form, then
\begin{eqnarray}
\omega={1\over k!} \omega_{i_1\dots i_k} dx^{i_1}\wedge\dots \wedge dx^{i_k}~,
\end{eqnarray}
and
\begin{eqnarray}
d\omega={1\over k!} \partial_{i_1} \omega_{i_2\dots i_{k+1}} dx^{i_1}\wedge\dots \wedge dx^{i_{k+1}}~,
\end{eqnarray}
leading to
\begin{eqnarray}
(d\omega)_{i_1\dots i_{k+1}}= (k+1) \partial_{[i_1} \omega_{i_2\dots i_{k+1}]}~.
\end{eqnarray}

Furthermore, we write
\begin{eqnarray}
\omega^2= \omega_{i_1\dots i_k} \omega^{i_1\dots i_k}~,~~~\omega^2_{i_1 i_{2}}=\omega_{i_1j_1\dots j_{k-1}} \omega_{i_2}{}^{j_1\dots j_{k-1}} \ .
\end{eqnarray}
Given a volume form $d\mathrm{vol}={1\over n!} \epsilon_{i_1\dots i_n} dx^{i_1}\wedge \dots \wedge dx^{i_n}$, the Hodge dual of $\omega$ is defined as
\begin{eqnarray}
*\omega\wedge\chi = (\chi, \omega) d\mathrm{vol}
\end{eqnarray}
where
\begin{eqnarray}
(\chi, \omega)={1\over k!} \chi_{i_1\dots i_k} \omega^{i_1\dots i_k}~.
\end{eqnarray}
So
\begin{eqnarray}
*\omega_{i_1\dots i_{n-k}}={1\over k!} \epsilon_{i_1\dots i_{n-k}}{}^{j_1\dots j_k} \omega_{j_1\dots j_k}~.
\end{eqnarray}
In particular the (anti) self-duality of the IIB 5-form field strengths is given by
\begin{eqnarray}
F_{M_1\dots M_5}=-{1\over 5!} \epsilon_{M_1\dots M_5}{}^{N_1\dots N_5} F_{N_1\dots N_5}~,
\end{eqnarray}
eg $F_{+ - {z 3 4}} = -F_{{5 6 7 8 9}} $.
For complex forms
\begin{eqnarray}
\parallel \omega\parallel^2 =\bar\omega_{i_1\dots i_k} \omega^{i_1\dots i_k}~.
\end{eqnarray}
It is well-known that for every form $\omega$, one can define a Clifford algebra element ${\slashed \omega}$ given by
\begin{eqnarray}
{\slashed\omega}=\omega_{i_1\dots i_k} \Gamma^{i_1\dots i_k}~,
\end{eqnarray}
where $\Gamma^i$, $i=1,\dots n$, are the Dirac gamma matrices. In addition we introduce the notation
\begin{eqnarray}
{\slashed\omega}_{i_1}= \omega_{i_1 i_2 \dots i_k} \Gamma^{i_2\dots i_k}~,~~~\Gamma\slashed{\omega}_{i_1}= \Gamma_{i_1}{}^{
i_2\dots i_{k+1}} \omega_{i_2\dots i_{k+1}}~.
\end{eqnarray}
The rest of our spinor conventions can be found in \cite{iibsystem}.

\section{$AdS_3$: Proof of the maximum principle}\label{clifads3}

In this appendix, we shall derive (\ref{ads3maxp}).  This involves extensive Clifford algebra manipulations and
the use of the field equations, in particular the scalar part of the Einstein equation on $M^7$. For this, we consider $\parallel\chi_+\parallel^2$, assume $\mathscr{D}^{(+)}\chi_+=0$, and evaluate $\nabla^2 \left\| \chi_+ \right\| ^2$
to find
\begin{eqnarray} \label{eq:first_expansion}
 \nabla^2 \left\| \chi_+ \right\| ^2 &=& 2 \left\| \mathbb{D}^{( + )} \chi_+ \right\| ^2 + \frac{1}{2} R^{(7)} \left\| \chi_+ \right\| ^2
 \cr
 && + \operatorname{Re}\operatorname{Re} \left\langle \chi_+, \left[ -4 {\Psi}^{( + ) i\dagger} - 2 \Psi^{( + ) i} - 2 \Gamma^{i j} \Psi^{( + )}_{j} - 14 q A^{-1} \Gamma^{z i} \mathbb{B}^{(+)} \right] {\nabla}_{i} \chi_+ \right\rangle
\cr
 && + \operatorname{Re} \left\langle \chi_+, \left[ -2 \left( {\Psi}^{( + ) i \dagger} + q A^{-1} {\mathbb{B}}^{(+)\dagger} \Gamma^{z i} \right) \left( \Psi^{( + )}_{i} + q A^{-1} \Gamma_{z i} \mathbb{B}^{(+)} \right) \right. \right.
 \cr
 && \left. \left. - 2 {\nabla}^{i} \Psi^{( + )}_{i} - 2 \Gamma^{i j} {\nabla}_{i} \Psi^{( + )}_{j} - 14{\nabla}_{i} \left( q A^{-1} \Gamma^{z i} \mathbb{B}^{(+)} \right) \right] \chi_+ \right\rangle ,
\end{eqnarray}
where
\begin{align}
 {\Psi}^{( + )\dagger}_{i} &= \frac{1}{2 A} \partial_{i} A + \frac{i}{2} Q_{i} - \frac{i}{4} \left( \slashed{\Gamma Y} \right) _i \Gamma_z - \frac{i}{2} \slashed{Y}_i \Gamma_z
 \nonumber \\
 & \qquad \qquad + \left( -\frac{1}{96} \left( \slashed{\Gamma H} \right) _i - \frac{9}{96} \slashed{H}_i + \frac{6}{96} \Phi \Gamma_{z i} \right) C *~,
 \nonumber \\
 {\mathbb{B}}^{(+)\dagger} &= -\frac{c}{2\ell} - \frac{1}{2} \slashed{\partial} A \, \Gamma_z + \frac{i A}{4} \slashed{Y} + \left( -\frac{A}{96} \slashed{H} \Gamma_z + \frac{18 A}{96} \Phi \right) C * .
\end{align}
Expanding out the third term, we find that
\begin{align}
 &  {\mkern-72mu}\operatorname{Re} \left\langle \chi_+, \left[ -4 {\Psi}^{( + ) i \dagger} - 2 \Psi^{( + ) i} - 2 \Gamma^{i j} \Psi^{( + )}_{j} - 14 q A^{-1} \Gamma^{z i} \mathbb{B}^{(+)} \right] {\nabla}_{i} \chi_+ \right\rangle
 \nonumber \\
 &= \operatorname{Re} \left\langle \chi_+, \left[ \frac{7 q c}{\ell} A^{-1} \Gamma^{z i} - (3 + 7 q) \partial^{i} \log A - (1 + 7 q) \left( \Gamma \slashed{\partial}\log A \right) ^i \right. \right.
  \nonumber \\
 &  - i Q^{i} + i \left( \slashed{\Gamma Q} \right) ^i + \frac{-2 + 14 q}{2} i \slashed{Y}^i \Gamma^z + \frac{-1 + 7 q}{2} i \left( \slashed{\Gamma Y} \right) ^i \Gamma^z
 \nonumber \\
 &  \left. \left. + \left( \frac{-6 + 14 q}{96} \left( \slashed{\Gamma H} \right) ^i + \frac{6 + 42 q}{96} \slashed{H}^i + \frac{60 - 252 q}{96} \Phi \Gamma^{z i} \right) C * \right] {\nabla}_{i} \chi_+ \right\rangle .
\end{align}
For $ q = \frac{1}{7} $, this term can be rewritten as
\begin{align} \nonumber
 & {\mkern-72mu} \operatorname{Re} \left\langle \chi_+, \left[ -4 {\Psi}^{( + ) i \dagger} - 2 \Psi^{( + ) i} - 2 \Gamma^{i j} \Psi^{( + )}_{j} - 14 q A^{-1} \Gamma^{z i} \mathbb{B}^{(+)} \right] {\nabla}_{i} \chi_+ \right\rangle
 \nonumber \\
 &= -3 \partial^{i}\log A \, {\nabla}_{i} \left\| \chi_+ \right\| ^2 + \operatorname{Re} \left\langle \chi_+, \mathcal{F} \Gamma^i {\nabla}_{i} \chi_+ \right\rangle
 \nonumber \\
 &=  -3 \partial^{i} \log A \, {\nabla}_{i} \left\| \chi_+ \right\| ^2 - \operatorname{Re} \left\langle \chi_+, \mathcal{F} \Gamma^i \left[ \Psi^{( + )}_{i} + \frac{1}{7 A} \Gamma_{z i} \mathbb{B}^{(+)} \right] \chi_+ \right\rangle ,
\end{align}
where
\begin{equation}
 \mathcal{F} = {c\over \ell} A^{-1} \Gamma^z + 2 \slashed{\partial}\log A - i \slashed{Q} + \left( \frac{1}{24} \slashed{H} + \frac{1}{4} \Phi \Gamma^z \right) \, C * .
\end{equation}
Combining the ${\cal{F}}$-term with the bilinear part of the fourth term in \eqref{eq:first_expansion}, we find that
\begin{align} \nonumber
 &  \operatorname{Re} \left\langle \chi_+, -2 \left( {\Psi}^{( + ) i \dagger} + \frac{1}{7 A} {{ \mathbb{B}}}^{(+) \dagger} \Gamma^{z i} + \frac{1}{2} \mathcal{F} \, \Gamma^i \right) \left( \Psi^{( + )}_{i} + \frac{1}{7 A} \Gamma_{z i} \mathbb{B}^{(+)} \right) \chi_+ \right\rangle
 \nonumber \\
 &= \operatorname{Re} \left\langle \chi_+, -2 \left[ \frac{3 c}{7 \ell A} \Gamma^{z i} + \frac{10}{7 A} \partial^{i} A - \frac{13}{14 A} \left( \Gamma \slashed{\partial} A \right) ^i + \frac{i}{2} \left( \slashed{\Gamma Q} \right) ^i - \frac{3 i}{7} \slashed{Y}^i \Gamma_z \right. \right.
 \nonumber \\
 & \qquad \qquad \qquad \qquad \left. \left. - \frac{2 i}{7} \left( \slashed{\Gamma Y} \right) ^i \Gamma^z + \left( -\frac{20}{7 \cdot 96} \left( \slashed{\Gamma H} \right) ^i - \frac{24}{7 \cdot 96} \slashed{H}^i + \frac{3}{14} \Phi \Gamma^{z i} \right) C * \right] \right.
 \nonumber \\
 & \qquad \qquad \left. \times \left[ -\frac{c}{14 \ell A} \Gamma_{z i} + \frac{4}{7 A} \partial_{i} A + \frac{1}{14 A} \left( \Gamma \slashed{\partial} A \right) _i - \frac{i}{2} Q_{i} - \frac{4 i}{7} \slashed{Y}_{i} \Gamma_z \right. \right.
 \nonumber \\
 & \qquad \qquad \qquad \qquad \left. \left. + \frac{3 i}{14} \left( \slashed{\Gamma Y} \right) _i \Gamma_z + \left( -\frac{8}{7 \cdot 96} \left( \slashed{\Gamma H} \right) _i + \frac{60}{7 \cdot 96} \slashed{H}_i - \frac{1}{28} \Phi \Gamma_{z i} \right) C * \right] \chi_+ \right\rangle
 \nonumber \\
 &= \operatorname{Re} \left\langle \chi_+, \left[ -\frac{3}{7 \ell^2 A^2} - \frac{17}{7 A^2} ( dA ) ^2 - \frac{8 i}{7} A^{-1} \partial_i A \slashed{Y}^i - \frac{4 i c}{7\ell A} \slashed{Y} - 2 Y^2 - \frac{1}{7} \slashed{Y}^2 \right. \right.
 \nonumber \\
 &  \indent - \frac{3}{28} \parallel \Phi \parallel ^2 + \frac{1}{84} \Phi \slashed{\overline{H}} \Gamma_z - \frac{11}{7 \cdot 288} \slashed{H} \slashed{\overline{H}} + \frac{1}{32} \slashed{H}^i \slashed{\overline{H}}_i
\nonumber \\
 &  \indent \left. \left. + \left( \frac{i}{12} Q_{i} \left( \slashed{\Gamma H} \right) ^i + \frac{c}{42\ell A} \slashed{H} \Gamma_z - \frac{1}{42 A} \partial_i A \left( \slashed{\Gamma H} \right) ^i + \frac{i}{14} \slashed{Y}^i \slashed{H}_i \Gamma_z - \frac{3 c}{7\ell A} \Phi \right) C * \right] \chi_+ \right\rangle \ .
\end{align}

We can use the field equations and Bianchi identities to rewrite the last line of \eqref{eq:first_expansion} as
\begin{align} \nonumber
 & {\mkern-72mu} \operatorname{Re} \left\langle \chi_+, \left[ -2 {\nabla}^{i} \Psi^{( + )}_{i} - 2 \Gamma^{i j} {\nabla}_{i} \Psi^{( + )}_{j} - {\nabla}_{i} \left( \frac{2}{A} \Gamma^{z i} \mathbb{B}^{(+)} \right) \right] \chi_+ \right\rangle
 \\ \nonumber
 &= \operatorname{Re} \left\langle \chi_+, \left( \frac{2}{A^2} ( dA ) ^2 - \frac{2}{A} {\nabla}^2 A + \frac{i}{2} \slashed{dQ} - 2 i {\nabla}_i \slashed{Y}^i \Gamma_z - \frac{1}{48} \slashed{dH} C * \right) \chi_+ \right\rangle
 \\ \nonumber
 &= \operatorname{Re} \left\langle \chi_+, \left[ \frac{4}{\ell^2 A^2} + \frac{6}{A^2} ( dA ) ^2 - 4 Y^2 - \frac{3}{4} \parallel \Phi \parallel ^2 + \xi_{i} \overline{\xi}_{j} \Gamma^{i j} \right. \right.
 \\ \nonumber
 & \qquad \qquad \qquad \qquad + \frac{1}{144} \slashed{H} \slashed{\overline{H}} - \frac{1}{16} \slashed{H}^i \slashed{\overline{H}}_i - \frac{1}{8} \slashed{H}^{i j} \slashed{\overline{H}}_{i j}
 \\
 & \qquad \qquad \qquad \qquad \left. \left. + \left( - \frac{i}{12} Q_{i} \left( \slashed{\Gamma H} \right) ^i + \frac{1}{12} \xi_{i} \overline{ \left( \slashed{\Gamma H} \right) } {}^i \right) C* \right] \chi_+ \right\rangle .
\end{align}

The second, fourth, and ${\cal{F}}$-term part of the third term  on the right side of equation \eqref{eq:first_expansion} thus sum to
\begin{align} \nonumber
 & \operatorname{Re} \left\langle \chi_+, \left[ \frac{4}{7\ell^2 A^2} + \frac{4}{7 A^2} ( dA ) ^2 - \frac{4 i c}{7\ell A} \slashed{Y} - \frac{8 i}{7} A^{-1} \partial_i A \slashed{Y}^i - \frac{1}{7} \slashed{Y}^2 + \frac{1}{7} \parallel \Phi \parallel ^2 \right. \right.
 \\ \nonumber
 & \qquad \qquad + \frac{1}{84} \Phi \slashed{\overline{H}} \Gamma_z + \parallel \xi \parallel ^2 + \xi_{i} \overline{\xi}_{j} \Gamma^{i j} + \frac{1}{7 \cdot 96} \slashed{H} \slashed{\overline{H}} - \frac{1}{32} \slashed{H}^i \slashed{\overline{H}}_i - \frac{1}{8} \slashed{H}^{i j} \slashed{\overline{H}}_{i j}
 \\ \nonumber
 & \qquad \qquad + \frac{1}{12} \parallel H \parallel ^2 + \left( \frac{c}{42 \ell A} \slashed{H} \Gamma^z - \frac{1}{42 A} \partial_{i} A \, \left( \slashed{\Gamma H} \right) ^i + \frac{1}{12} \xi_{i} \overline{ \left( \slashed{\Gamma H} \right) } {}^i \right.
 \\
 & \qquad \qquad \qquad \qquad \qquad \qquad \left. \left. \left. + \frac{i}{14} \slashed{Y}^i \slashed{H}_i \Gamma_z - \frac{3 c}{7\ell  A} \Phi \right) C * \right] \chi_+ \right\rangle \ .
\end{align}

Noting that
\begin{align}
 \left\| \mathbb{B}^{(+)} \chi_+ \right\| ^2 &= \left\langle \chi_+, {{\mathbb{B}}}^{(+)\dagger} \mathbb{B}^{(+)} \chi_+ \right\rangle
 \nonumber \\
 &= \left\langle \chi_+, \left[ \frac{1}{4\ell^2} + \frac{1}{4} ( dA ) ^2 - \frac{i A}{2} \slashed{Y}^{i} \partial_{i} A \, \Gamma_z - \frac{i A c}{4\ell} \slashed{Y} - \frac{A^2}{16} \slashed{Y}^2 \right. \right.
  \nonumber \\
 & \qquad \qquad - \frac{A^2}{96^2} \slashed{H} \slashed{\overline{H}} + \frac{9 A^2}{256} \parallel \Phi \parallel ^2 - \frac{A^2}{256} \Phi \slashed{\overline{H}} \Gamma_z
 \nonumber \\
 & \qquad \qquad \left. \left. + \left( \frac{A c}{96 \ell} \slashed{H} \Gamma^z - \frac{A}{96} \partial_{i} A \left( \slashed{\Gamma H} \right) ^i + \frac{i A^2}{32} \slashed{Y}_i \slashed{H}^i \Gamma_z - \frac{3 A c}{16\ell } \Phi \right) C * \right] \chi_+ \right\rangle
 \nonumber \\
 \left\| \mathcal{A}^{( + )} \chi_+ \right\| ^2 &= \operatorname{Re} \left\langle \chi, \left[ \parallel \xi \parallel ^2 + \xi_{i} \overline{\xi}_{j} \Gamma^{i j} + \frac{1}{576} \slashed{H} \slashed{\overline{H}} - \frac{1}{32} \slashed{H}^i \slashed{\overline{H}}_i \right. \right.
\nonumber \\
 & \qquad \qquad \qquad \qquad - \frac{1}{8} \slashed{H}^{i j} \slashed{\overline{H}}_{i j} + \frac{1}{12} \parallel H \parallel ^2 + \frac{1}{16} \parallel \Phi \parallel ^2
 \nonumber \\
 & \qquad \qquad \qquad \qquad \left. \left. + \frac{1}{48} \Phi \slashed{\overline{H}} \Gamma_z + \frac{1}{12} \xi_{i} \overline{ \left( \slashed{\Gamma H} \right) } {}^i C * \right] \chi \right\rangle
\end{align}
we can now write equation \eqref{eq:first_expansion} as  (\ref{ads3maxp}).
We also remark that the values of
$q$ for $AdS_n$ backgrounds are given by
$q={n-2 \over 10-n}$.

\section{$AdS_4$: Proof of the maximum principle}\label{clifads4}

To prove  (\ref{ads4maxpp}), we assume that $\mathscr{D}^{(+)}\chi_+=0$ and evaluate
\begin{align} \label{eq:first_expansion4}
 {\nabla}^2 \left\| \chi_+ \right\| ^2 &= 2 \left\| \mathbb{D}^{(+)} \chi_+ \right\| ^2 + \frac{1}{2} R^{(6)} \left\| \chi_+ \right\| ^2
  \nonumber \\
 & \qquad \qquad + \operatorname{Re} \left\langle \chi_+, \left[ -4 {\Psi}^{( + ) i \dagger} - 2 \Psi^{( + ) i} - 2 \Gamma^{i j} \Psi^{( + )}_{j} \right. \right.
 \nonumber \\
 & \qquad \qquad \qquad \qquad \left. \left. - 12 \frac{q}{A} \Gamma^{z i} \mathbb{B}^{(+)} \right] \nabla_{i} \chi_+ \right\rangle
 \nonumber \\
 & \qquad \qquad +\operatorname{Re} \left\langle \chi_+, \left[ -2 \left( {\Psi}^{( + ) i \dagger} + \frac{q}{A} {\mathbb{B}}^{(+)\dagger} \Gamma^{z i} \right) \left( \Psi^{( + )}_{i} + \frac{q}{A} \Gamma_{z i} \mathbb{B}^{(+)} \right) \right. \right.
 \nonumber \\
 & \qquad \qquad \qquad \qquad \left. \left. - 2 \nabla^{i} \Psi^{( + )}_{i} - 2 \Gamma^{i j} \nabla_{i} \Psi^{( + )}_{j} - 12 \nabla_{i} \left( \frac{q}{A} \Gamma^{z i} \mathbb{B}^{(+)} \right) \right] \chi_+ \right\rangle ,
\end{align}
where
\begin{align}
 {\Psi}^{( + )\dagger}_{i} &= \frac{1}{2 A} \partial_{i} A + \frac{i}{2} Q_{i} - \frac{i}{2} Y_{i} \Gamma_{z x} - \frac{i}{2} \left( \slashed{\Gamma Y} \right) _{i} \Gamma_{z x} + \left( -\frac{1}{96} \left( \slashed{\Gamma H} \right) _{i} - \frac{9}{96} \slashed{H}_{i} \right) C *~,
 \nonumber \\
 {\mathbb{B}}^{(+)\dagger} &= -\frac{c}{2\ell} - \frac{1}{2} \slashed{\partial}^{i} A \, \Gamma_z - \frac{i A}{2} \slashed{Y} \Gamma_x - \frac{A}{96} \slashed{H} \Gamma^z C * .
\end{align}

Expanding the third term in (\ref{eq:first_expansion4}), we find that
\begin{align}
 & {\mkern-72mu} \operatorname{Re} \left\langle \chi_+, \left[ -4 {\Psi}^{( + ) i\dagger} - 2 \Psi^{( + ) i} - 2 \Gamma^{i j} \Psi^{( + )}_{j} - 12 \frac{q}{A} \Gamma^{z i} \mathbb{B}^{(+)} \right] \nabla_{i} \chi_+ \right\rangle
  \nonumber \\
 &= \operatorname{Re} \left\langle \chi_+, \left[ \frac{6 q c}{\ell A} \Gamma^{z i} - \frac{3 + 6 q}{A} \partial^{i} A - \frac{1 + 6 q}{A} \left( \Gamma \slashed{\partial} A \right) ^{i} - i Q^{i} + i \left( \slashed{\Gamma Q} \right) ^{i} \right. \right.
 \nonumber \\
 & \qquad \qquad + i \left( -2 + 6 q \right) Y^{i} \Gamma^{z x} + i \left( -2 + 6 q \right) \left( \slashed{\Gamma Y} \right) ^{i} \Gamma^{z x}
 \nonumber \\
 & \qquad \qquad \left. \left. + \left( \frac{-8 + 12 q}{96} \left( \slashed{\Gamma H} \right) ^{i} + \frac{36 q}{96} \slashed{H}^{i} \right) C * \right] \nabla_{i} \chi_+ \right\rangle .
\end{align}
For $ q = \frac{1}{3} $, this can be written as
\begin{align} \nonumber
 & {\mkern-72mu} \operatorname{Re} \left\langle \chi_+, \left[ -4 {\Psi}^{( + ) i\dagger} - 2 \Psi^{( + ) i} - 2 \Gamma^{i j} \Psi^{( + )}_{j} - 12 \frac{q}{A} \Gamma^{z i} \mathbb{B}^{(+)} \right] \nabla_{i} \chi_+ \right\rangle
 \nonumber \\
 &= -\frac{4}{A} \partial^{i} A \, \nabla_{i} \left\| \chi_+ \right\| ^2 + \operatorname{Re} \left\langle \chi_+, \mathcal{F} \Gamma^i \nabla_{i} \chi_+ \right\rangle
 \nonumber \\
 &=  -\frac{4}{A} \partial^{i} A \, \nabla_{i} \left\| \chi_+ \right\| ^2 - \operatorname{Re} \left\langle \chi_+, \mathcal{F} \Gamma^i \left[ \Psi^{( + )}_{i} + \frac{1}{3A} \Gamma_{z i} \mathbb{B}^{(+)} \right] \chi_+ \right\rangle ,
\end{align}
where
\begin{equation}
 \mathcal{F} = \frac{2 c}{\ell A} \Gamma^z + \frac{3}{A} \slashed{\partial} A - i \slashed{Q} + \frac{1}{24} \slashed{H} \, C * .
\end{equation}
Combining the ${\cal{F}}$-term with  the bilinear part of the fourth term in \eqref{eq:first_expansion4}, we find that
\begin{align} \nonumber
 &  \operatorname{Re} \left\langle \chi_+, -2 \left( {\Psi}^{( + ) i \dagger} + \frac{1}{3 A} {\mathbb{B}}^{(+)\dagger} \Gamma^{z i} + \frac{1}{2} \mathcal{F} \, \Gamma^i \right) \left( \Psi^{( + )}_{i} + \frac{1}{3 A} \Gamma_{z i} {\mathbb{B}}^{(+)} \right) \chi_+ \right\rangle
\nonumber \\
 &= \operatorname{Re} \left\langle \chi_+, -2 \left[ \frac{5 c}{6 \ell A} \Gamma^{z i} + \frac{11}{6 A} \partial^{i} A - \frac{4}{3 A} \left( \Gamma \slashed{\partial} A \right) ^i + \frac{i}{2} \left( \slashed{\Gamma Q} \right) ^i - \frac{i}{3} Y^{i} \Gamma_{z x} \right. \right.
  \nonumber \\
 & \qquad \qquad \qquad \qquad \left. \left. - \frac{2 i}{3} \left( \slashed{\Gamma Y} \right) ^i \Gamma^{z x} + \left( -\frac{1}{36} \left( \slashed{\Gamma H} \right) ^i - \frac{1}{24} \slashed{H}^i \right) C * \right] \right.
  \nonumber \\
 & \qquad \qquad \left. \times \left[ -\frac{c}{6 \ell A} \Gamma_{z i} + \frac{2}{3 A} \partial_{i} A + \frac{1}{6 A} \left( \Gamma \slashed{\partial} A \right) _i - \frac{i}{2} Q_{i} - \frac{2 i}{3} Y_{i} \Gamma_{z x} \right. \right.
  \nonumber \\
 & \qquad \qquad \qquad \qquad \left. \left. + \frac{i}{3} \left( \slashed{\Gamma Y} \right) _i \Gamma_{z x} + \left( -\frac{1}{72} \left( \slashed{\Gamma H} \right) _i + \frac{1}{12} \slashed{H}_i \right) C * \right] \chi_+ \right\rangle
 \nonumber \\
 &= \operatorname{Re} \left\langle \chi_+, \left[ -\frac{5}{3\ell^2 A^2} - \frac{14}{3 A^2} ( dA ) ^2 - \frac{8}{3} Y^{i} \partial_{i} A \, \Gamma_{z x} + \frac{8 i c}{3\ell A} \slashed{Y} \Gamma_x - \frac{8 i}{3 A} Y^2 \right. \right.
 \nonumber \\
 & \qquad \qquad \left. \left. + \left( \frac{i}{12} Q_{i} \left( \slashed{\Gamma H} \right) ^i + \frac{c}{18\ell A} \slashed{H} \Gamma^z + \frac{i}{6} Y_i \slashed{H}^i \Gamma_{z x} C * - \frac{1}{18 A} \partial_{i} A \, \left( \slashed{\Gamma H} \right) ^i \right) C * \right] \chi_+ \right\rangle \ .
\nonumber \\
\end{align}

We can use the field equations and Bianchi identities to rewrite the last line of \eqref{eq:first_expansion4} as
\begin{align} \nonumber
 & {\mkern-72mu} \operatorname{Re} \left\langle \chi_+, \left[ -2 \nabla^{i} \Psi^{( + )}_{i} - 2 \Gamma^{i j} \nabla_{i} \Psi^{( + )}_{j} - \nabla_{i} \left( \frac{4}{A} \Gamma^{z i} \mathbb{B}^{(+)} \right) \right] \chi_+ \right\rangle
 \\ \nonumber
 &= \operatorname{Re} \left\langle \chi_+, \left( \frac{3}{A^2} ( dA ) ^2 - \frac{3}{A} \nabla^2 A + \frac{i}{2} \slashed{dQ} - 2 i \nabla^i Y_i \Gamma_{z x} - \frac{1}{48} \slashed{dH} C * \right) \chi_+ \right\rangle
 \\ \nonumber
 &= \operatorname{Re} \left\langle \chi_+, \left[ \frac{9}{\ell^2 A^2} + \frac{12}{A^2} ( dA ) ^2 - 12 Y^2 - \frac{1}{16} \parallel H \parallel ^2 + \xi_{{i}} \overline{\xi}_{{j}} \Gamma^{i j} \right. \right.
 \\ \nonumber
 & \qquad \qquad + \frac{1}{144} H_{{i_1} {i_2} {i_3}} \overline{H}_{{j_1} {j_2} {j_3}} \Gamma^{i_1 i_2 i_3 j_1 j_2 j_3}
 \\
 & \qquad \qquad \left. \left. + \left( - \frac{i}{12} Q_{i} \left( \slashed{\Gamma H} \right) ^i + \frac{1}{12} \xi_{i} \overline{ \left( \slashed{\Gamma H} \right) } {}^i \right) C* \right] \chi_+ \right\rangle .
\end{align}
The second, fourth and ${\cal{F}}$-term part of the third term on the right side of equation \eqref{eq:first_expansion4} thus sum to
\begin{align}
 & \operatorname{Re} \left\langle \chi_+, \left[ \frac{4}{3 \ell^2 A^2} + \frac{4}{3 A^2} ( dA ) ^2 - \frac{8 i}{3} Y^i \partial_i A \, \Gamma_{z x} + \frac{8 i c}{3\ell A} \slashed{Y} \Gamma_x + \frac{4}{3} Y^2 \right. \right.
  \nonumber \\
 & \qquad \qquad \qquad \qquad + \parallel \xi \parallel ^2 + \xi_{i} \overline{\xi}_{j} \Gamma^{i j} - \frac{1}{864} \slashed{H} \slashed{\overline{H}} - \frac{1}{32} \slashed{H}^i \slashed{\overline{H}}_i - \frac{1}{8} \slashed{H}^{i j} \slashed{\overline{H}}_{i j} + \frac{1}{12} \parallel H \parallel ^2
  \nonumber \\
 & \qquad \qquad \qquad \qquad \left. \left. + \left( \frac{c}{18\ell A} \slashed{H} \Gamma^z - \frac{1}{18 A} \partial_{i} A \, \left( \slashed{\Gamma H} \right) ^i + \frac{i}{6} Y_i \slashed{H}^i \Gamma_{z x} C * \right) C * + \frac{1}{12} \xi_{i} \overline{ \left( \slashed{\Gamma H} \right) } {}^i \right] \chi_+ \right\rangle~.
\end{align}

Noting that
\begin{align}
 \left\| \mathbb{B}^{(+)} \chi_+ \right\| ^2 &= \left\langle \chi_+, \left[ \frac{1}{4\ell^2} + \frac{1}{4} ( dA ) ^2 + \frac{i A}{2} Y^{i} \partial_{i} A \, \Gamma_{z x} + \frac{i A c}{2\ell} \slashed{Y} \Gamma_x + \frac{A^2}{4} Y^2 \right. \right.
 \nonumber \\
 & \qquad \qquad \left. \left. - \frac{A^2}{96^2} \slashed{H} \slashed{\overline{H}} + \left( \frac{A c}{96\ell} \slashed{H} \Gamma^z - \frac{A}{96} \partial_{i} A \left( \slashed{\Gamma H} \right) ^i + \frac{i A^2}{32} Y^{i} \slashed{H}_i \Gamma^{z x} \right) C * \right] \chi_+ \right\rangle
 \nonumber \\
 \left\| \mathcal{A} \chi_+ \right\| ^2 &= \operatorname{Re} \left\langle \chi, \left[ \parallel \xi \parallel ^2 + \xi_{i} \overline{\xi}_{j} \Gamma^{i j} - \frac{1}{576} \slashed{\overline{H}} \slashed{H} + \frac{1}{12} \xi_{i} \overline{ \left( \slashed{\Gamma H} \right) } {}^i C * \right] \chi \right\rangle
\end{align}
we can now write equation \eqref{eq:first_expansion4} as  (\ref{ads4maxpp}).

\section{$AdS_5$: Proof of the maximum principle}\label{clifads5}
To prove the formula (\ref{ads5maxpp}), we evaluate
\begin{align} \label{eq:first_expansion5}
 \nabla^2 \left\| \chi_+ \right\| ^2 &= 2 \left\| \mathbb{D}^{(+)} \chi_+ \right\| ^2 + \frac{1}{2} R^{(5)} \left\| \chi_+ \right\| ^2
 \nonumber \\
 & \qquad \qquad + \operatorname{Re} \left\langle \chi_+, \left[ -4 {\Psi}^{( + ) {i}\dagger} - 2 \Psi^{( + ) {i}} - 2 \Gamma^{i j} \Psi^{( + )}_{j} \right. \right.
 \nonumber \\
 & \qquad \qquad \qquad \qquad \left. \left. - 10 \frac{q}{A} \Gamma^{z i} \mathbb{B}^{(+)} \right] \nabla_{{i}} \chi_+ \right\rangle
 \nonumber \\
 & \qquad \qquad + \operatorname{Re} \left\langle \chi_+, \left[ -2 \left( {\Psi}^{( + ) {i}\dagger} + \frac{q}{A} {\mathbb{B}}^{(+)\dagger} \Gamma^{z i} \right) \left( \Psi^{( + )}_{{i}} + \frac{q}{A} \Gamma_{z i} \mathbb{B}^{(+)} \right) \right. \right.
 \nonumber \\
 & \qquad \qquad \qquad \qquad \left. \left. - 2 \nabla^{{i}} \Psi^{( + )}_{{i}} - 2 \Gamma^{i j} \nabla_{{i}} \Psi^{( + )}_{j} - 10 \nabla_{{i}} \left( \frac{q}{A} \Gamma^{z i} \mathbb{B}^{(+)} \right) \right] \chi_+ \right\rangle ,
\end{align}
where
\begin{align}
 {\Psi}^{( + )\dagger}_{{i}} &= \frac{1}{2 A} \partial_{{i}} A + \frac{i}{2} Q_{{i}} + \frac{i}{2} Y \Gamma_{z x y i} + \left( -\frac{1}{96} \left( \slashed{\Gamma H} \right) _i - \frac{9}{96} \slashed{H}_i \right) C *~,
 \nonumber \\
 {\mathbb{B}}^{(+)\dagger} &= -\frac{c}{2\ell} + \frac{1}{2} \partial^{{i}} A \, \Gamma_{z i} + \frac{i A}{2} Y \Gamma_{x y} - \frac{A}{96} \slashed{H} \Gamma^z C * .
\end{align}

Expanding the third term in (\ref{eq:first_expansion5}), we find that
\begin{align}
 & {\mkern-72mu} \operatorname{Re} \left\langle \chi_+, \left[ -4 {\Psi}^{( + ) {i}\dagger} - 2 \Psi^{( + ) {i}} - 2 \Gamma^{i j} \Psi^{( + )}_{j} - 10 \frac{q}{A} \Gamma^{z i} \mathbb{B}^{(+)} \right] \nabla_{{i}} \chi_+ \right\rangle
  \nonumber \\
 &= \operatorname{Re} \left\langle \chi_+, \left[ \frac{5 q c}{\ell A} \Gamma^{z i} - \frac{3 + 5 q}{A} \partial^{{i}} A - \frac{1 + 5 q}{A} \partial_{j} A \, \Gamma^{i j} \right. \right.
 \nonumber \\
 & \qquad \qquad -i Q^{{i}} + i Q_{j} \Gamma^{i j} + i \left( 3 - 5 q \right) Y \Gamma^{z x y i}
  \nonumber \\
  & \qquad \qquad \left. \left. + \left( \frac{-10 + 10 q}{96} \left( \slashed{\Gamma H} \right) ^i + \frac{-6 + 30 q}{96} \slashed{H}^i \right) C * \right] \nabla_{{i}} \chi_+ \right\rangle .
\end{align}
For $ q = \frac{3}{5} $, the above expression can be rewritten as
\begin{align} \nonumber
 & {\mkern-72mu} \operatorname{Re} \left\langle \chi_+, \left[ -4 {\Psi}^{( + ) {i}\dagger} - 2 \Psi^{( + ) {i}} - 2 \Gamma^{i j} \Psi^{( + )}_{j} - 10 \frac{q}{A} \Gamma^{z i} \mathbb{B}^{(+)} \right] \nabla_{{i}} \chi_+ \right\rangle
\nonumber \\
 &= -\frac{5}{A} \partial^{{i}} A \left\| \chi_+ \right\| ^2 + \operatorname{Re} \left\langle \chi_+, \mathcal{F} \Gamma^i \nabla_{{i}} \chi_+ \right\rangle
\nonumber  \\
 &= -\frac{5}{A} \partial^{{i}} A \left\| \chi_+ \right\| ^2 - \operatorname{Re} \left\langle \chi_+, \mathcal{F} \Gamma^i \left[ \Psi^{( + )}_{{i}} + \frac{3}{5 A} \Gamma_{z i} \Xi_+ \right] \chi_+ \right\rangle ,
\end{align}
where
\begin{equation}
 \mathcal{F} = \frac{3 c}{\ell A} \Gamma^z + \frac{4}{A} \slashed{\partial} A - i \slashed{Q} + \frac{1}{24} \slashed{H} C *~.
\end{equation}
Combining the ${\cal{F}}$-term with the bilinear part of the fourth term of equation \eqref{eq:first_expansion5}, we find that
\begin{align} \nonumber
 & {\mkern-72mu} \operatorname{Re} \left\langle \chi_+, -2 \left( {\Psi}^{( + ) {i}\dagger} + \frac{3}{5 A} {\mathbb{B}}^{(+)\dagger} \Gamma^{z i} + \frac{1}{2} \mathcal{F} \, \Gamma^i \right) \left( \Psi^{( + )}_{{i}} + \frac{3}{5 A} \Gamma_{z i} \mathbb{B}^{(+)} \right) \chi_+ \right\rangle
\nonumber \\
 &= \operatorname{Re} \left\langle \chi_+, -2 \left[ \frac{6 c}{5\ell A} \Gamma^{z i} + \frac{11}{5 A} \partial^{{i}} A - \frac{17}{10 A} \partial_{j} A \, \Gamma^{i j} + \frac{4 i}{5} Y \Gamma^{z x y i} \right. \right.
 \nonumber \\
 & \qquad \qquad \qquad \qquad \left. + \frac{i}{2} Q_{j} \Gamma^{i j} + \left( -\frac{1}{40} \left( \slashed{\Gamma H} \right) ^i - \frac{1}{20} \slashed{H}^i \right) C * \right]
\nonumber \\
 & \qquad \qquad \times \left[ -\frac{3 c}{10\ell A} \Gamma_{z i} + \frac{4}{5 A} \partial_{{i}} A + \frac{3}{10 A} \partial^{j} A \, \Gamma_{i j} - \frac{i}{5} Y \Gamma_{z x y i} \right.
  \nonumber \\
 & \qquad \qquad \qquad \qquad \left. \left. - \frac{i}{2} Q_{{i}} + \left( -\frac{1}{60} \left( \slashed{\Gamma H} \right) _i + \frac{3}{40} \slashed{H}_i \right) C * \right] \chi_+ \right\rangle
 \nonumber \\
 &= \operatorname{Re} \left\langle \chi_+, \left[ -\frac{18}{5\ell^2 A^2} - \frac{38}{5 A^2} ( dA ) ^2 - \frac{8}{5} Y^2 - \frac{24 i c}{5\ell A} Y \Gamma_{x y} \right. \right.
 \nonumber \\
 & \qquad \qquad - 2 \left( -\frac{1}{40} \left( \slashed{\Gamma H} \right) ^i - \frac{1}{20} \slashed{H}^i \right) \left( -\frac{1}{60} \overline{ \left( \slashed{\Gamma H} \right) } {}_i + \frac{3}{40} \slashed{\overline{H}}_i \right)
 \nonumber \\
 &  \qquad \qquad \left. \left. + \left( \frac{c}{10 \ell A} \slashed{H} \Gamma_z + \frac{1}{16 A} \partial_i A \left( \slashed{\Gamma H} \right) ^i + \frac{i}{12} Q_i \left( \slashed{\Gamma H} \right) ^i \right) C * \right] \chi_+ \right\rangle .
\end{align}
We can use the field equations and  Bianchi identities to rewrite the last line of \eqref{eq:first_expansion} as
\begin{align} \nonumber
 & {\mkern-72mu} \operatorname{Re} \left\langle \chi_+, \left[ -2 \nabla^{i} \Psi^{( + )}_{i} - 2 \Gamma^{i j} \nabla_{i} \Psi^{( + )}_{j} - \nabla_{i} \left( \frac{6}{A} \Gamma^{z i} \mathbb{B}^{(+)} \right) \right] \chi_+ \right\rangle
 \nonumber \\
 &= \operatorname{Re} \left\langle \chi_+, \left( \frac{4}{A^2} ( dA ) ^2 - \frac{4}{A} \nabla^2 A + \frac{i}{2} \slashed{dQ} - \frac{1}{48} \slashed{dH} C * \right) \chi_+ \right\rangle
 \nonumber \\
 &= \operatorname{Re} \left\langle \chi_+, \left[ \frac{16}{\ell^2 A^2} + \frac{20}{A^2} ( dA ) ^2 - 16 Y^2 - \frac{1}{12} \parallel H \parallel ^2 + \xi_i \overline{\xi}_j \Gamma^{i j} \right. \right.
 \nonumber \\
 & \qquad \qquad \qquad \qquad \left. \left. + \left( -\frac{i}{12} Q_{i} \left( \slashed{\Gamma H} \right) ^i + \frac{1}{12} \xi_{i} \overline{ \left( \slashed{\Gamma H} \right) } {}^i \right) C * \right] \chi_+ \right\rangle .
\end{align}
The second, fourth and ${\cal{F}}$-term part of the third term on the right side of equation \eqref{eq:first_expansion5} thus sum to
\begin{align} \nonumber
 &\operatorname{Re} \left\langle \chi_+, \left[ \frac{12}{5\ell^2 A^2} + \frac{12}{5 A^2} ( dA ) ^2 + \frac{12}{5} Y^2 - \frac{24 i c}{5\ell A} Y \Gamma_{x y} + \parallel \xi \parallel ^2 + \xi_i \overline{\xi}_j \Gamma^{i j} \right. \right.
 \\ \nonumber
 & \qquad \qquad \qquad \qquad - \frac{1}{40} H^i{}_{j_1 j_2} \overline{H}_{i k_1 k_2} \Gamma^{j_1 j_2 k_1 k_2} - \frac{1}{80} H^{i_1 i_2}{}_j \overline{H}_{i_1 j_2 k} \Gamma^{j k} + \frac{1}{60} \parallel H \parallel ^2
 \\
 & \qquad \qquad \qquad \qquad \left. \left. + \left( \frac{c}{10 \ell A} \slashed{H} \Gamma_z - \frac{1}{10 A} \partial_i A \left( \slashed{\Gamma H} \right) ^i + \frac{1}{12} \xi_{i} \overline{ \left( \slashed{\Gamma H} \right) } {}^i \right) C * \right] \chi_+ \right\rangle \ .
\end{align}
Noting that
\begin{align}
 \left\| \mathbb{B}^{(+)} \sigma_+ \right\| ^2 &= \left\langle \chi_+, {\mathbb{B}}^{(+)\dagger} \mathbb{B}^{(+)} \chi_+ \right\rangle
\nonumber \\
 &= \left\langle \chi_+, \left[ \frac{1}{4\ell^2 } + \frac{1}{4} ( dA ) ^2 + \frac{i A}{2} Y^{i} \partial_{i} A \, \Gamma_{z x} - \frac{i A c}{2\ell } Y \Gamma_{x y} + \frac{A^2}{4} Y^2 \right. \right.
 \nonumber \\
 & \qquad \qquad - \frac{A^2}{32^2} H^i{}_{j_2 j_3} \overline{H}_{i k_2 k_3} \Gamma^{j_2 j_3 k_2 k_3} + \frac{A^2}{2 \cdot 16^2} H^{i_1 i_2}{}_j \overline{H}_{i_1 i_2 k} \Gamma^{j k}
 \nonumber \\
 & \qquad \qquad \left. \left. + \frac{A^2}{16 \cdot 96} \parallel H \parallel ^2 + \left( \frac{A c}{96\ell} \slashed{H} \Gamma^z - \frac{A}{96} \partial_{i} A \left( \slashed{\Gamma H} \right) ^i \right) C * \right] \chi_+ \right\rangle
 \nonumber \\
 \left\| \mathcal{A} \chi_+ \right\| ^2 &= \operatorname{Re} \left\langle \chi, \left[ \parallel \xi \parallel ^2 + \xi_{i} \overline{\xi}_{j} \Gamma^{i j} - \frac{1}{64} H^i{}_{j_1 j_2} \overline{H}_{i k_1 k_2} \Gamma^{j_1 j_2 k_1 k_2} \right. \right.
 \nonumber \\
 & \qquad \qquad \left. \left. - \frac{1}{32} H^{i_1 i_2}{}_j \overline{H}_{i_1 i_2 k} \Gamma^{j k} + \frac{1}{96} \parallel H \parallel ^2 + \frac{1}{12} \xi_{i} \overline{ \left( \slashed{\Gamma H} \right) } {}^i C * \right] \chi \right\rangle
\end{align}
we can now write equation \eqref{eq:first_expansion5} as (\ref{ads5maxpp}).

\section{$AdS_6$: Proof of the maximum principle}\label{clifads6}

To prove (\ref{ads6maxpp}), we evaluate
\begin{align} \label{eq:s_+_first_expansion6}
 \nabla^2 \left\| \chi \right\| ^2 &= 2 \left\| \mathbb{D}^{(+)} \chi \right\| ^2 + \frac{1}{2} R^{(4)} \left\| \chi \right\| ^2
  \nonumber \\
 & \qquad \qquad + \operatorname{Re} \left\langle \chi, \left[ -4 {\Psi}^{( + ) i \dagger} - 2 \Psi^{( + ) i} - 2 \Gamma^{i j} \Psi^{( + )}_{j} - 8 \frac{q}{A} \Gamma^{z i} \mathbb{B}^{(+)} \right] \nabla_{i} \chi \right\rangle
  \nonumber \\
 & \qquad \qquad + \operatorname{Re} \left\langle \chi, \left[ -2 \left( {\Psi}^{( + ) i \dagger} + \frac{q}{A} {\mathbb{B}}^{(+)\dagger} \Gamma^{z i} \right) \left( \Psi^{( + )}_{i} + \frac{q}{A} \Gamma_{z i} \mathbb{B}^{(+)} \right) \right. \right.
  \nonumber \\
 & \qquad \qquad \qquad \qquad \left. \left. - 2 \nabla^{i} \Psi^{( + )}_{i} - 2 \Gamma^{i j} \nabla_{i} \Psi^{( + )}_{j} - 8 \nabla_{i} \left( \frac{q}{A} \Gamma^{z i} \mathbb{B}^{(+)} \right) \right] \chi \right\rangle ,
\end{align}
where
\begin{align}
 {\Psi}^{( + )\dagger}_{i} &= \frac{1}{2 A} \partial_{i} A + \frac{i}{2} Q_{i} + \left( -\frac{1}{96} \left( \slashed{\Gamma H} \right) _i - \frac{9}{96} \slashed{H}_i \right) C *
\nonumber  \\
 {\mathbb{B}}^{(+)\dagger} &= -\frac{c}{2\ell} + \frac{1}{2} \partial_{i} A \, \Gamma^{z i} - \frac{A}{96} \slashed{H} \Gamma^z C * \ .
\end{align}

Using the fact that $ \operatorname{Re} \left\langle \phi, \Gamma^{i j} \phi \right\rangle = \operatorname{Re} \left\langle \phi, \Gamma^{i j} C * \phi \right\rangle = 0 $, we can expand the third term in (\ref{eq:s_+_first_expansion6}) as
\begin{align} \nonumber
 & {\mkern-72mu} \qquad \operatorname{Re} \left\langle \chi, \left[ -4 {\Psi}^{( + ) i\dagger} - 2 \Psi^{( + ) i} - 2 \Gamma^{i j} \Psi^{( + )}_{j} - 8 \frac{q}{A} \Gamma^{z i} \mathbb{B}^{(+)} \right] \nabla_{i} \chi \right\rangle
\nonumber \\
 &= \operatorname{Re} \left\langle \chi, \left[ \frac{4 c q}{\ell A} \Gamma^{z i} - \frac{3 + 4 q}{A} \partial^{i} A - \frac{1 + 4 q}{A} \partial_{j} A \, \Gamma^{i j} - i Q^{i} + i Q_{j} \Gamma^{i j} \right. \right.
 \nonumber \\
 & \qquad \qquad \qquad \qquad \qquad \qquad \left. \left. + \left( \frac{-12 + 8 q}{96} \left( \slashed{\Gamma H} \right) ^i + \frac{-12 + 24 q}{96} \slashed{H}^i \right) C * \right] \nabla_{i} \chi \right\rangle .
\end{align}
For $q=1$, the above term can be rewritten as
\begin{align} \nonumber
 & {\mkern-72mu} \qquad \operatorname{Re} \left\langle \chi, \left[ -4 {\Psi}^{( + ) i\dagger} - 2 \Psi^{( + ) i} - 2 \Gamma^{i j} \Psi^{( + )}_{j} - 8 \frac{q}{A} \Gamma^{z i} \mathbb{B}^{(+)} \right] \nabla_{i} \chi \right\rangle
 \\
 &= -\frac{6}{A} \partial^{i} A \nabla_{i} \left\| \chi \right\| ^2 - \operatorname{Re} \left\langle \chi, \mathcal{F} \, \Gamma^i \left[ \Psi^{( + )}_{i} + \frac{1}{A} \Gamma_{z i} \mathbb{B}^{(+)} \right] \chi \right\rangle
\end{align}
where
\begin{equation}
 \mathcal{F} = \frac{4 c}{\ell A} \Gamma^z + \frac{5}{A} \slashed{\partial} A - i \slashed{Q} + \frac{1}{24} \slashed{H} \, C * .
\end{equation}
Combining the ${\cal{F}}$-term with the bilinear part of the fourth term in \eqref{eq:s_+_first_expansion6}, we find that
\begin{align} \nonumber
 & \operatorname{Re} \left\langle \chi, -2 \left( {\Psi}^{( + ) i\dagger} + \frac{1}{A} {\mathbb{B}}^{(+)\dagger} \Gamma^{z i} + \frac{1}{2} \mathcal{F} \, \Gamma^i \right) \left( \Psi^{( + )}_{i} + \frac{1}{A} \Gamma_{z i} \mathbb{B}^{(+)} \right) \chi \right\rangle
 \nonumber \\
 &= \operatorname{Re} \left\langle \chi, -2 \left[ \frac{3 c}{2\ell A} \Gamma^{z i} + \frac{5}{2 A} \partial^{i} A - \frac{2}{A} \partial_{j} A \, \Gamma^{i j} + \frac{i}{2} Q_{j} \Gamma^{i j} + \left( -\frac{2}{96} \left( \slashed{\Gamma H} \right) ^i - \frac{6}{96} \slashed{H}^i \right) C * \right] \right.
 \nonumber \\
 & \indent \times \left. \left[ -\frac{c}{2\ell A} \Gamma_{z i} + \frac{1}{A} \partial_{i} A + \frac{1}{2 A} \partial^{j} A \, \Gamma_{i j} - \frac{i}{2} Q_{i} + \left( -\frac{2}{96} \left( \slashed{\Gamma H} \right) _i + \frac{6}{96} \slashed{H}^i \right) C * \right] \chi \right\rangle
 \nonumber \\
 &= \operatorname{Re} \left\langle \chi, \left[ -\frac{6}{\ell^2 A^2} - \frac{11}{A^2} ( dA ) ^2 - 2 \left( -\frac{2}{96} \left( \slashed{\Gamma H} \right) ^i - \frac{6}{96} \slashed{H}^i \right) \left( -\frac{2}{96} \overline{ \left( \slashed{\Gamma H} \right) } {}_i + \frac{6}{96} \slashed{\overline{H}}_i \right) \right. \right.
 \nonumber \\
 &  \qquad \qquad \left. \left. + \left( \frac{16 c}{96\ell A} \slashed{H} \Gamma^z - \frac{16}{96 A} \partial_{i} A \left( \slashed{\Gamma H} \right) ^i + \frac{i}{12} Q_{i} \left( \slashed{\Gamma H} \right) ^i \right) C * \right] \chi \right\rangle \ .
\end{align}
Next, we use the field equations and Bianchi identities to expand the derivatives in the fourth term on the right side of equation \eqref{eq:s_+_first_expansion6} as
\begin{align} \nonumber
 & {\mkern-72mu} \operatorname{Re} \left\langle \chi, \left[ -2 \nabla^{i} \Psi^{( + )}_{i} - 2 \Gamma^{i j} \nabla_{i} \Psi^{( + )}_{j} - 8 \nabla_{i} \left( \frac{1}{A} \Gamma^{z i} \mathbb{B}^{(+)} \right) \right] \chi \right\rangle
 \nonumber \\
 &= \operatorname{Re} \left\langle \chi, \left[ \frac{5}{A^2} ( dA ) ^2 - \frac{5}{A} \nabla^2 A + \frac{i}{2} \slashed{dQ} - \frac{1}{48} \slashed{dH} \, C * \right] \chi \right\rangle
 \nonumber \\
 &= \operatorname{Re} \left\langle \chi, \left[ \frac{25}{\ell^2 A^2} + \frac{30}{A^2} ( dA ) ^2 - \frac{5}{48} \parallel H \parallel ^2 + \xi_{i} \overline{\xi}_{j} \Gamma^{i j} \right. \right.
 \nonumber \\
 & \qquad \qquad \qquad \qquad \left. \left. + \left( -\frac{i}{12} Q_{i} \left( \slashed{\Gamma H} \right) ^i + \frac{1}{12} \xi_{i} \overline{ \left( \slashed{\Gamma H} \right) } {}^i \right) C * \right] \chi \right\rangle .
\end{align}
The second, fourth and ${\cal{F}}$-term part of the third term on the right side of equation \eqref{eq:s_+_first_expansion6} thus sum to
\begin{align}
 & \operatorname{Re} \left\langle \chi, \left[ \frac{4}{\ell^2 A^2} + \frac{4}{A^2} ( dA ) ^2 + \frac{1}{48} \parallel H \parallel ^2 + \parallel \xi \parallel ^2 + \xi_i \overline{\xi}_j \Gamma^{i j} \right. \right.
 \nonumber \\
 & \left. \left. \qquad \qquad \qquad \qquad + \left( \frac{16 c}{96\ell A} \slashed{H} \Gamma^z - \frac{6}{96 A} \partial_{i} A \left( \slashed{\Gamma H} \right) ^i  + \frac{1}{12} \xi_{i} \overline{ \left( \slashed{\Gamma H} \right) } {}^i \right) C * \right] \chi \right\rangle \ .
\end{align}
Noting that
\begin{align}
 \left\| \mathbb{B}^{(+)} \chi \right\| ^2 &= \operatorname{Re} \left\langle \chi, {\mathbb{B}}^{(+)\dagger} \mathbb{B}^{(+)} \chi \right\rangle
 \nonumber \\
 &= \operatorname{Re} \left\langle \chi, \left[ \frac{1}{4\ell^2 } + \frac{1}{4} ( dA ) ^2 + \frac{3 A^2}{16 \cdot 96} H^{i j}{}_k \overline{H}_{i j \ell} \Gamma^{k \ell} + \frac{A^2}{16 \cdot 96} \parallel H \parallel ^2 \right. \right.
 \nonumber \\
 & \qquad \qquad \qquad \qquad \left. \left. + \left( \frac{A c}{96\ell} \slashed{H} \Gamma^z - \frac{A}{96} \partial_{i} A \, \left( \slashed{\Gamma H} \right) ^i \right) C * \right] \chi \right\rangle
 \nonumber \\
 \left\| \mathcal{A} \chi \right\| ^2 &= \operatorname{Re} \left\langle \chi, \mathcal{A}^\dagger \mathcal{A} \chi \right\rangle
 \nonumber \\
 &= \operatorname{Re} \left\langle \chi, \left[ \parallel \xi \parallel ^2 + \xi_{i} \overline{\xi}_{j} \Gamma^{i j} - \frac{3}{96} H^{i j}{}_k \overline{H}_{i j \ell} \Gamma^{k \ell} + \frac{1}{96} \parallel H \parallel ^2 + \frac{1}{12} \xi_{i} \overline{ \left( \slashed{\Gamma H} \right) } {}^i C * \right] \chi \right\rangle
\end{align}
we can now write equation \eqref{eq:s_+_first_expansion6} as (\ref{ads6maxpp}).

\end{document}